\documentclass[aps,groupedaddress]{revtex4-1}

\usepackage{url} 
\usepackage{graphicx}
\usepackage{epstopdf}
\usepackage{subfig,float}
\usepackage{amsmath,amssymb}
\usepackage{natbib}
\usepackage{xcolor}
\usepackage{dcolumn}%
\usepackage{bm}%
\newcommand{\lb}{\label}
\newcommand{\lra}[1]{\langle #1 \rangle }

\newcommand{\fltr}{\overline}
\newcommand{\grad}{{\mbox{\boldmath $\nabla$}}}

\newcommand{\fu}{\fltr\bu}

\newcommand{\btau}{{\mbox{\boldmath $\tau$}}}
\newcommand{\br}{{\bf r}}

\newcommand{\bk}{{\bf k}}
\newcommand{\bq}{{\bf q}}

\newcommand{\bu}{{\bf u}}

\newcommand{\bx}{{\bf x}}

\def\be{\begin{equation}}
\def\ee{\end{equation}}

\def\RMV#1{{\textcolor{gray}{  }}}

\begin{document}


\title{ On the local energy flux of turbulent flows }

\author{Alexandros Alexakis}
\email[]{alexakis@phys.ens.fr}
\affiliation{Laboratoire de Physique de l’Ecole normale supérieure, ENS, Université PSL, CNRS, Sorbonne Université, Université de Paris, F-75005 Paris, France}

\author{Sergio Chibbaro}
\email[]{sergio.chibbaro@upmc.fr}
\affiliation{Sorbonne Universit\'e, CNRS, UMR 7190, Institut Jean Le Rond d'Alembert, F-75005 Paris, France}


\date{\today}

\begin{abstract}

We investigate the local energy flux rate $\Pi_\ell(\bx)$ towards small scales in isotropic turbulent flows using  direct numerical simulations and applying different low-pass filters. Two different filters are examined, a sharp Fourier filter and a Gaussian filter. The probability density function (pdf) of the local energy flux is calculated for the different filters and for different filtering scales. It is shown that the local energy flux is a largely fluctuating quantity taking both negative and positive values and this is more pronounced for the sharp filter. The variance, the skewness and the kurtosis of these fluctuations are shown to increase as the filtering scale is decreased. Furthermore we calculate the joint pdf of $\Pi_\ell(\bx)$ with the local filtered strain rate $S_\ell$ and the enstrophy $\Omega_\ell$. The flux shows a good correlation with the strain but not with the enstrophy.  It is shown that its conditional mean value scales like $\lra{\Pi_\ell}_S \propto \ell^2 S_\ell^{3} $
in support to the Smagorinsky eddy viscosity model. Nonetheless strong fluctuations 
exist around this value that also need to be modeled. We discuss the implications of our results for subgrid scale models, and propose new modelling directions.

\end{abstract}


\maketitle

\section{Introduction}
Turbulent flows are the fundamental basis of many engineering applications \cite{Mon_75,Pope_turbulent}, geophysics \cite{thorpe2007introduction,gill2016atmosphere}, and astrophysics \cite{biskamp2003magnetohydrodynamic} among others. To correctly capture the behavior of these complex problems with direct numerical simulation is in principle possible but out of question for the foreseeable time to come. This is because performing such simulations requires that all scales from the largest (for instance of the order of thousands km for the atmosphere) to the smallest disipative scales (of the order of cm for the atmosphere) need to be resolved. This leads to an enormous number of degrees of freedom whose evolution needs to be followed. In practice, a modelling approach is needed, where only the scales of interest are kept while the remaining scales are omitted and their effect on the former scales needs to be estimated. Because of the complexity of these flows, including complex geometry and several physical mechanisms at play, the most common way to arrive at such estimates is through semi-empirical approaches \cite{pope2000turbulent, Mon_75, wilcox1998turbulence}. In the last decades, an approach which has gained in popularity is the Large-Eddy-Simulation (LES) \cite{germano1992turbulence, lesieur1996new, piomelli1999large, Pope_turbulent, meneveau2000scale}. In this framework, a low-pass filter is formally applied to equations, and only a part of the degrees of freedom (the large-scale motion) is directly solved, while the remaining scales are filtered out and their effect on the resolved scales is modelled by additional terms in the dynamical equations. Unfortunately, there is no separation of scales between large and small scales in turbulent flows to perform an asymptotic expansion, and therefore it is not possible to close the problem in a rigorous way \cite{Ten_90,Fri_95,bohr2005dynamical}. It is therefore necessary to build up phenomenological closures based on our understanding of the physics of small-scale turbulence. 
Such phenomenological closures however need to be thoroughly tested with real data. For this goal, direct numerical simulations (DNS) appear as the most valuable tool to get insights permitting the assessment of present models and their improvement \cite{reynolds1990potential, Moi_98,pope2000turbulent, Fox2003}

For any such modeling attempt, it is key to find a good compromise between including an accurate physics and keeping simple and computationally efficient the structure of the model. With this in mind, in this work we have focused on the analysis of the most important element in turbulence dynamics, that is the energy-flux underlying the cascade \cite{alexakis2018cascades}. We investigate therefore how the transfer of energy to the subgrid scales can be modeled and on what observable of the resolved scales it should be based on. In order to address this question we perform a high-resolution numerical simulation and apply a scale-by-scale analysis based on the original approach by Germano \cite{germano1992turbulence}. This allows us to simultaneously measure the effect of the small filtered scales on the un-filtered scales as in \cite{borue1998local,chen2003joint, Chen:2006p1741, Eyink:2006p1379, chen2006kelvin} but at the same time to associate it with different observables of the large scales. 
%
To identify such observables we look at the gradients of the flow that have been useful in giving insights on the cascade mechanisms \cite{Ten_90,Chevillard:2008p4119,chevillard2011lagrangian}. A similar approach was used in the important work by Borue \& Orszag \cite{borue1998local} at low Reynolds number, where many insights are already given. 

More specifically, in this work we report on a comprehensive analysis of the flux in relation to the total strain and vorticity. This is achieved by calculating the probability density function (pdf) of the local energy flux for different filters and for different filtering scales as well as the joint pdf of the flux with two observables the strain and enstrophy of the filtered flow. This allows us for the first time to extract correlations between the flux and the two observables.  Our results strongly support the use of the Smagorinsky model\cite{smagorinsky1963general} but also emphasises its drawbacks. We conclude by discussing the implications of our results to modeling and also propose a possible  new subgrid scale model.

\section{Theoretic background}  
\subsection{Definitions} 
We begin by considering the incompressible Navier-Stokes equations describing the evolution of the velocity field $\bf u$ of an incompressible unit density fluid given by
\begin{eqnarray}
\ {\partial {\bf u}\over\partial t}+{\bf u}\cdot\nabla{\bf u} &=&
   -\nabla p + 
{\nu}\nabla^2\ {\bf u +f} \\
\nabla &\cdot& {\bf u}=0,
  \label{momentum}
\end{eqnarray}
where $p$ is the pressure, $\nu$ is the viscosity and $\bf f$ an external body force.
The flow is contained in a cube of side $2\pi$ and periodic boundary conditions are assumed.

To introduce the notion of different scales in the flow we use a filtering or coarse-graining approach \cite{germano1992turbulence}, where the dynamic velocity field $\bu$ is spatially (low-pass) filtered over a scale $\ell$ to obtained a filtered value $\fltr{\bu}_\ell(\bx)$. The filtering procedure is given by  
\be 
\fltr{\bu}_\ell(\bx) = \int d^3 r\, G_\ell(\br) \bu(\bx+\br) 
\label{eq:filter}
\ee
where $G_\ell$ is a smooth filtering function, spatially localized and such that $G_\ell (\vec r) = \ell^{-3}G(\vec r/\ell)$ where the function $G$ satisfies
$\int d\vec r \ G(\vec r)=1$, and $\int d\vec r \ \vert \vec r \vert ^2 G(\vec r) = \mathcal{O}(1)$. 
By applying the filtering to Navier-Stokes equations we obtain the coarse-grained dynamics 
\be 
\partial_{t} \overline{\bu}_\ell 
+  (\overline{\bu}_\ell \cdot \grad)\overline{\bu}_\ell = -\grad\overline{p}_{\ell} 
-\grad\cdot\btau_\ell
+\nu\nabla^{2}\overline{\bu}_\ell.
\label{eq:u-eq-ell} 
\ee
Here  $\btau_\ell$ is the 
{\it subscale stress tensor} (or momentum flux) which describes the force exerted 
on scales larger than $\ell$ by fluctuations at scales smaller than $\ell$. It is given by:
\be 
(\btau_\ell)_{i,j} = 
\overline{(u_i u_j)}_\ell -
(\overline{u}_\ell)_i (\overline{u}_\ell)_j 
\ee 
The corresponding pointwise kinetic energy budget reads
\begin{eqnarray}
\partial_t \left(\frac{1}{2}|\fu|^2\right) &+& \partial_j\left[ \left(\frac{1}{2}|\fu|^2 +\overline{p} \right)) \overline{u}_j  + \tau_{ij}\overline{u}_i
- \nu\partial_j\left(\frac{1}{2}|\fu|^{2}\right)\right] \\ \nonumber
&=& -\Pi_\ell  
 - \nu|\grad\fu|^2.
\lb{kinetic-large}
\end{eqnarray}
where we have dropped the $\ell$ subscript whenever unambiguous for the sake of clarity, and 
\be \Pi_\ell(\bx) \equiv -(\partial_{j}\overline{u}_{i})\tau_{ij}. 
\lb{kinetic-flux}
\ee
is the sub-grid scale (SGS) energy flux. This term is key since it represents the space-local transfer of energy among large and small scales across the scale $\ell$. In the case of direct energy cascade, the flux is known to be positive in average.

In homogeneous flows, an efficient way to implement the filter is to use its Fourier transform 
\be \hat{G}_q (\bk) = \int G_\ell({\bf x}) e^{i\bf k\cdot x}  d {\bx} \ee
where $q=1/\ell$ is the filtering wavenumber. 
In this work we consider two types of filters.
First we consider a Gaussian kernel 
\be \hat{G}_q(\bk)=\exp\left[-\frac{k^2}{2q^2}\right]. 
\label{gauss}\ee 
For an infinite domain this filter reduces to the Gaussian filter in real space $G_\ell(r) =\exp(-\frac{1}{2}r^2/\ell^2) /(2\pi \ell^{2})^{3/2}$ 
We note that this filtering is not a projection and in general
$\overline{(\overline{\bu_\ell})_\ell}\ne \overline{\bu_\ell} $. The second filter we are going to use is a sharp spectral filter such that 
\be
\overline{\bu}_\ell(\bx,t)=\sum_{\vert \bk \vert < q}\hat{\bu}(\bk,t) e^{i\bk \cdot \bx } .
\label{sharp}
\ee
This filtering is  a projector  $\left(\overline{(\overline{\bu_\ell})_\ell} = \overline{\bu_\ell} \right)$
and is based on a Galerkin truncation for all wavenumbers larger than the given cutoff $q=1/\ell$. 
%
This filtering is related to the classical definition of the energy flux $\Pi(q)$ given by
\be 
\Pi(q) = \left\langle
\overline{\bu}_\ell (\bu \cdot \nabla) \bu
\right\rangle 
\label{fluxGLB}
\ee 
where the angular brackets stand for spatial average
and the eq. (\ref{sharp}) has been used for $\overline{\bu}_\ell$.
Furthermore, when the sharp filtering is used the relation 
\be 
\langle \Pi_\ell(\bx) \rangle = \Pi(q)
\ee 
holds.


\subsection{Modeling} %
\label{modeling}      %

Given that in a LES only scales larger than $\ell$ 
are resolved it is desirable to model the subscale stress tensor $\btau_{i,j}$ based on the resolved scales and their geometric structure so that a closed system of equations is obtained. The simplest choice is to relate $\btau_{i,j}$ to the velocity gradient tensor of the filtered field $\grad{\overline{\bu}}_\ell$. It can be decomposed into its symmetric and anti-symmetric parts as $\grad{\overline{\bu}}_\ell=\overline{\bf S}_\ell+\overline{\bf \Omega}_\ell$, with 
\be \overline{\bf S}_\ell=(\grad{\overline{\bu}}_\ell+\grad{\overline{\bu}}_\ell^T)/2 \quad \mathrm{and} \quad \overline{\bf \Omega}_\ell=(\grad{\overline{\bu}}_\ell-\grad{\overline{\bu}}_\ell^T)/2.\ee 
The symmetric part is related to the strain, whereas the anti-symmetric to the vorticity. It is worth recalling that these quantities are related in average $\lra{{\vert\overline{\bf S}_\ell\vert^2}} =\lra{{\vert\overline{ \bf \Omega}_\ell\vert^2}}$ \cite{Fri_95}, but the local properties are not.
Using these tensors , the sub-scale energy flux is defined as $\Pi_\ell=\overline{\bf S}_\ell : \btau_\ell$, so that it is clear that only the symmetric part of resolved gradients enter directly in the definition of the flux. Yet, the dependence of the subscale stress $\btau$ on the strain and vorticity is not known {\it a priori}.

Several attempts to model the subscale stress tensor by $\overline{\bf S}_\ell$ and $\overline{\bf \Omega}_\ell$ have been made\cite{meneveau2000scale,pope2000turbulent}. The simplest models use the norms of the tensors $S_\ell^2={\vert\overline{\bf S}_\ell\vert^2}$ and $\Omega_\ell^2={\vert\overline{ \bf \Omega}_\ell\vert^2}$.
From these the most popular model is given by the Smagorinsky model \cite{smagorinsky1963general} where $\bf \btau$ is modeled as
\be 
{\btau}_{i,j} \approx - C_s^2 \ell^2 S_\ell \overline{\bf S}_{i,j}
\label{SmagoT}
\ee 
where $C_s$ is an order one non-dimensional number and
we use the symbol $\approx$ to indicate that the relation above is a model and is not an exact result. This expression gives the following estimate for the local energy
flux
\be
\Pi_\ell \approx C_s^2 \ell^2 {S}_\ell^3.
\label{Smago}
\ee 

Other models take into account $\overline{ \bf \Omega}_\ell$ as well. Indeed, approximating the sub-scale stress with its extreme local expression, that is as a function of the resolved scale, the nonlinear Clark model is obtained \cite{meneveau2000scale}:
\begin{equation}
\tau_\ell(\bu, \bu)\approx 
\dfrac{1}{3}C_2 \ell^2 \left(\overline{\bf S}_\ell^2+
\overline{\bf \Omega}_\ell^2+\overline{\bf \Omega}_\ell\overline{\bf S}_\ell-\overline{\bf S}_\ell\overline{\bf \Omega}_\ell \right )~,
\label{clark}
\end{equation}
where  both strain and vorticity participate in the dynamics \cite{misra1997vortex,borue1998local}.
For this model the formula for the flux is:
\begin{equation}
    \Pi_\ell\approx \dfrac{1}{3}C_2 \ell^2 [-\mathrm{Tr}(\overline{\bf S}_\ell^3)+3\mathrm{Tr}(\overline{\bf S}_\ell \overline{\bf \Omega}_\ell^2)]~,
\end{equation}
which shows that the local behaviour of the flux depends on a term related to pure strain and on the term linked to vortex-streching \cite{Ten_90}.
Generally speaking, the {\it local} dynamics of strain and vortex stretching can be quite independent \cite{tsinober2009informal}, and therefore a complete picture of the cascade requires both. Nonetheless, it is well known that for a homogeneous average there is the following kinematic relation\cite{betchov1956inequality}: $\lra{-\mathrm{Tr}(\overline{\bf S}_\ell^3)}=\lra{9 \mathrm{Tr}(\overline{\bf S}_\ell \overline{\bf \Omega}_\ell^2)}$, such that the mean flux can be related to the sole vortex-stretching term (or the strain skewness). Furthermore, the similarity of the statistics of these two terms was observed previously in a different context\cite{tsinober2000vortex}.
This suggests that within the purpose of modelling the cascade flux, the use of the sole strain may be justified. 
Nonetheless, all mentioned models are based on assumptions that can not be proven from basic principles. Therefore it is required confirmation from numerical simulations and experiments. 
One thus has to compare the results of direct numerical simulations (DNS) with different LES models \cite{vreman1997large, meneveau2000scale}, or alternatively one can use DNS to test directly the assumptions used by the models, the so-called {\it a priori} approach\cite{piomelli1999large}.  This later choice is what we are trying to do in the following sections.


\section{Results}                   
\subsection{The flow}   

We apply the formalism described in the previous section on the results of a direct numerical simulation of the Navier Stokes equations given by eq. (\ref{momentum}). The forcing was chosen so that there is a constant injection of energy at the Fourier modes with $|\bf k| \le k_f = 2$ and is explicitly given 
in terms of its Fourier components as
\be 
{ \hat{\bf f}_\bk} = 
\epsilon \sum_{|\bf k| \le 2}
\frac{\hat{\bu}_\bk }{\sum_{{|\bf k|} \le 2} |\hat{\bu}_\bk|^2}\, + \,
i \sum_{{|\bf k|}\le 2} \omega_\bk \hat{\bu}_\bk.
\ee 
where $\epsilon$ is the constant in time energy injection rate
that here we fix to $\epsilon=1$. The frequencies $\omega_\bk$ are chosen randomly in order to de-correlate the forced modes. 
\begin{figure*}
\begin{center}
\includegraphics[width=0.49\textwidth]{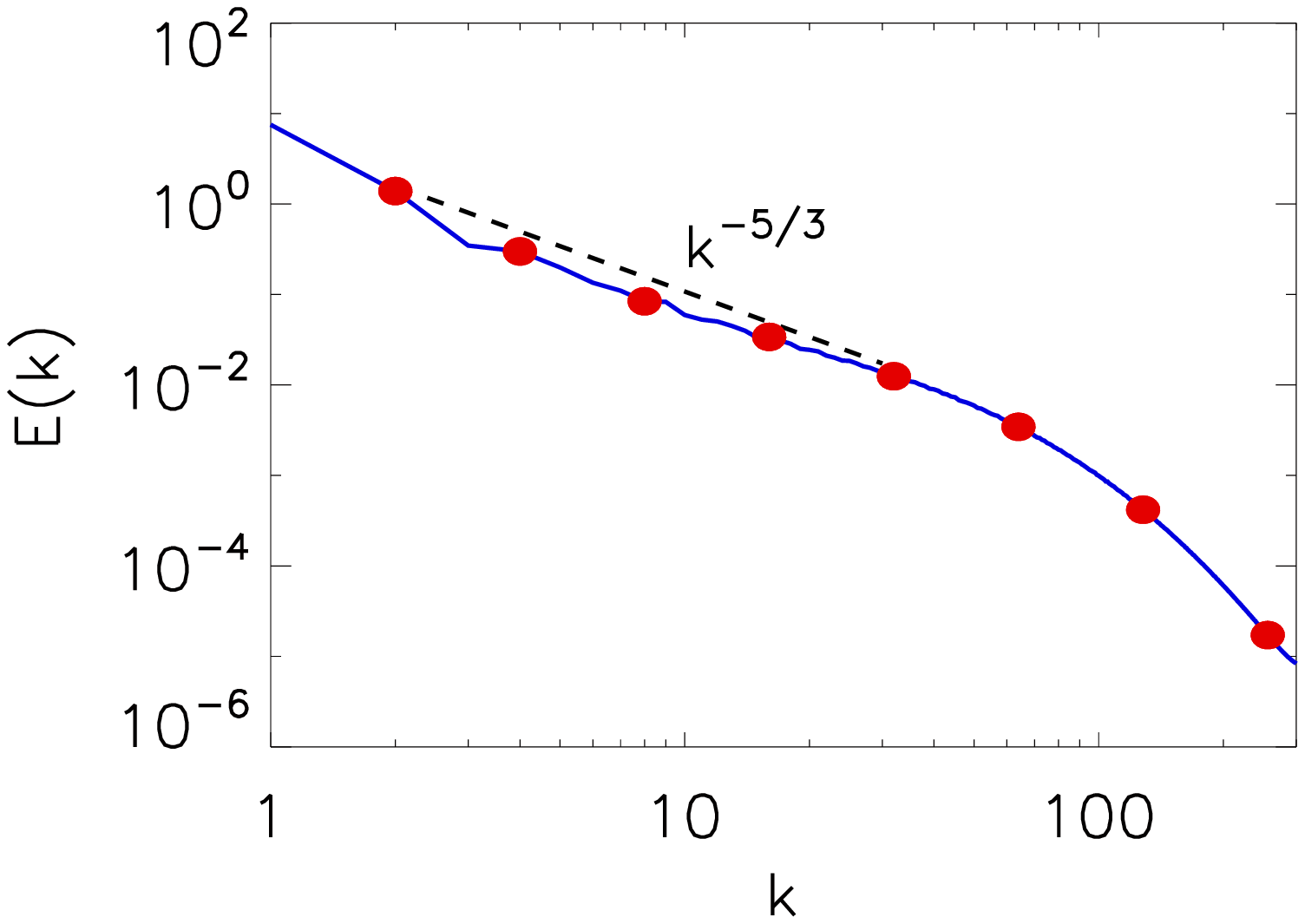}
\includegraphics[width=0.49\textwidth]{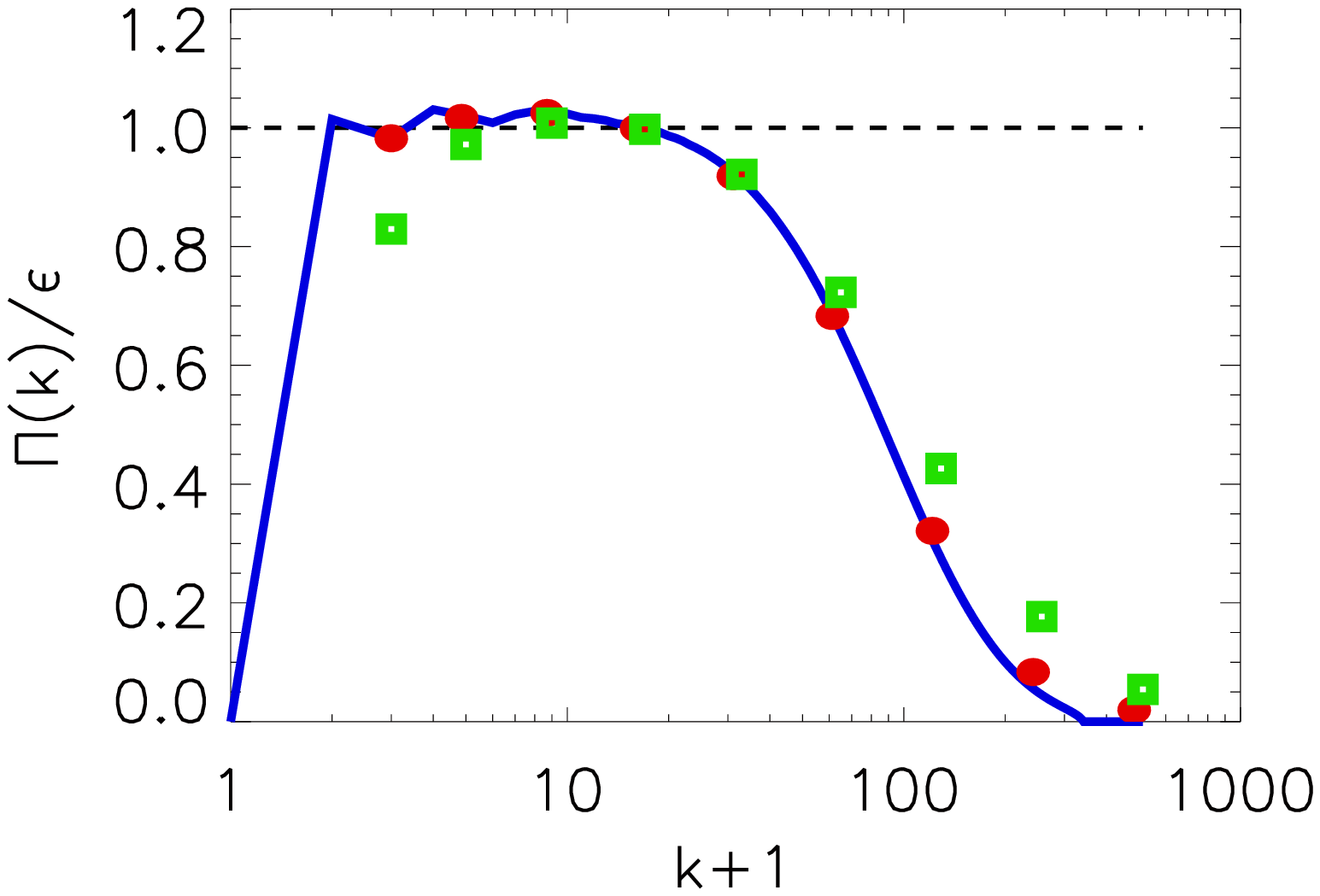}
\end{center}
\caption{Left: Energy spectrum for the simulation analysed. The forcing is at $k_f=1$. A dashed line with slope $-5/3$ is added for reference. The red dots indicate the wavenumbers at which the filtering was applied in following sections. Right: Average energy flux $\lra{\Pi_\ell}$ normalised with the mean rate of dissipation obtained without filtering (DNS) and with the sharp filter (red dots) and the Gaussian filter (green diamonds).}
\label{fig1}
\end{figure*} 
The flow was simulated using the pseudospectral code {\sc ghost} \cite{mininni2011hybrid} with a second order Runge-Kutta for time advancement and using the 2/3 rule for removing de-aliasing errors. The simulations were carried out with $\nu=0.0005$ on a $1024^3$ grid leading in each direction to a maximum wave number $k_{max}=N/3\simeq 341.$
The Reynolds number $Re=\epsilon^{1/3}k_1^{-4/3}/\nu$ achieved with this resolution was $Re=2000$ where $k_1=1$ is the smallest nonzero wavenumber in the domain. After a short transient the flow reaches a steady state where the energy dissipation rate matches the energy injection rate and the flow shows all characteristics of a classical isotropic turbulent flow.  In Fig. \ref{fig1}, we show the energy spectrum defined as 
\[ 
E(k) = \frac{1}{2}
\sum_{k-1<|\bq|\le k} |\hat{\bu}_q|^2 
\]
where $\hat{\bu}_q$ is the velocity Fourier mode.
The spectrum shows a standard behaviour, with a reasonable inertial range following a Kolmogorov power-law scaling  $E(k)\propto k^{-5/3}$ until around $k=50$. The red dots indicate the wavenumbers where filtering was applied that is examined in the next sections.
In the right panel, we show the energy flux that is almost constant in the inertial range. 
The flux marked by a solid blue line was calculated directly in Fourier space as is typically done in pseudospectral codes using eq. (\ref{fluxGLB}). The red circles indicate the space averaged local flux for the sharp filter of eq. (\ref{sharp}) at the different filtering wavenumbers $q$ that are going to be examined in the remaining sections of this work. With green squares, the space averaged local flux for the Gaussian filter of eq. (\ref{gauss}) is also shown for the same wavenumbers. As expected, the sharp projector perfectly overlaps on the solid blue line, while the mean energy flux obtained from the Gaussian filter does not match exactly, notably at small scales \cite{buzzicotti2018effect}. The two results are considerably close however in the inertial range.

\begin{figure*}
\begin{center}
\includegraphics[width=0.49\textwidth]{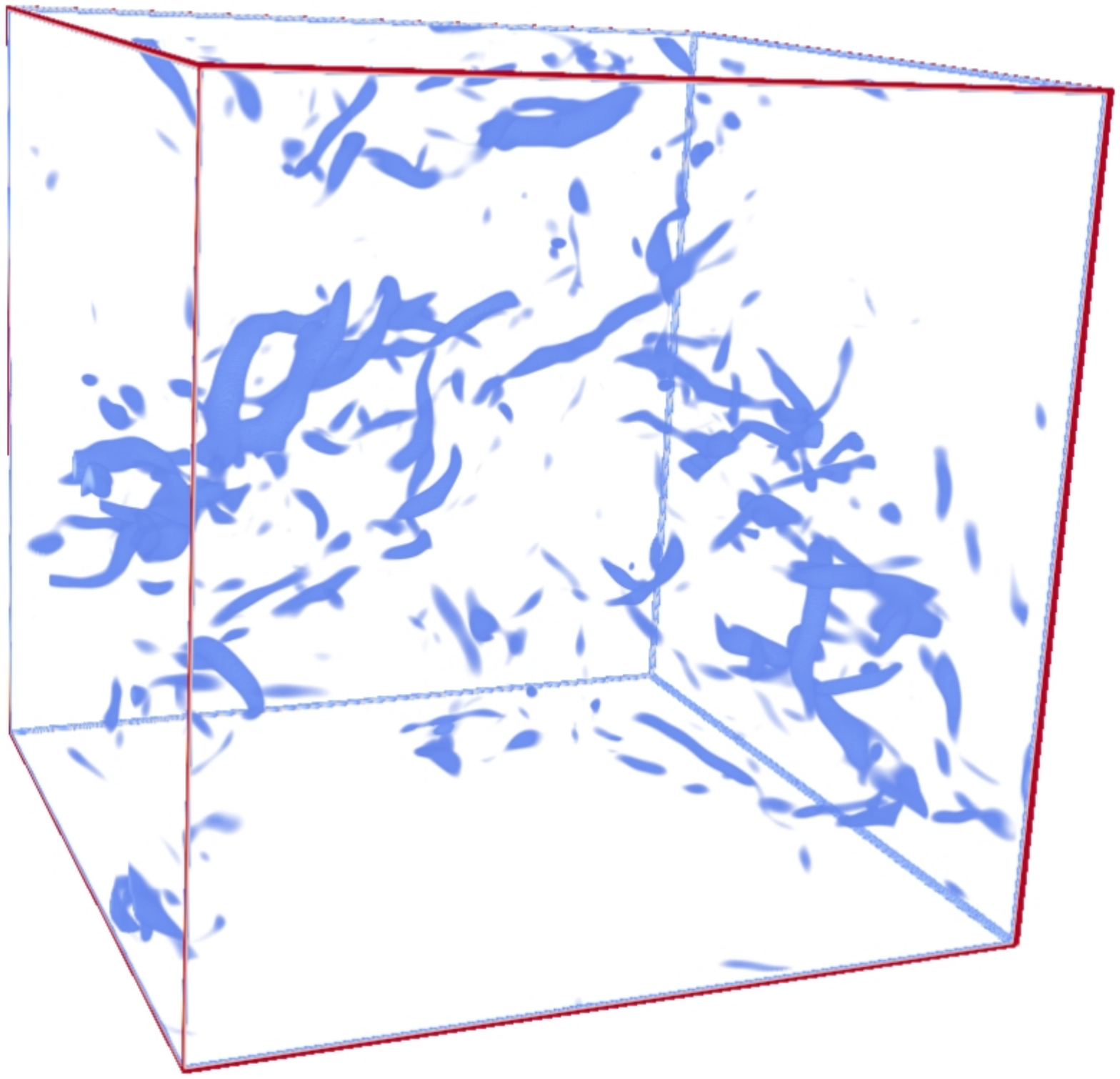}
\includegraphics[width=0.49\textwidth]{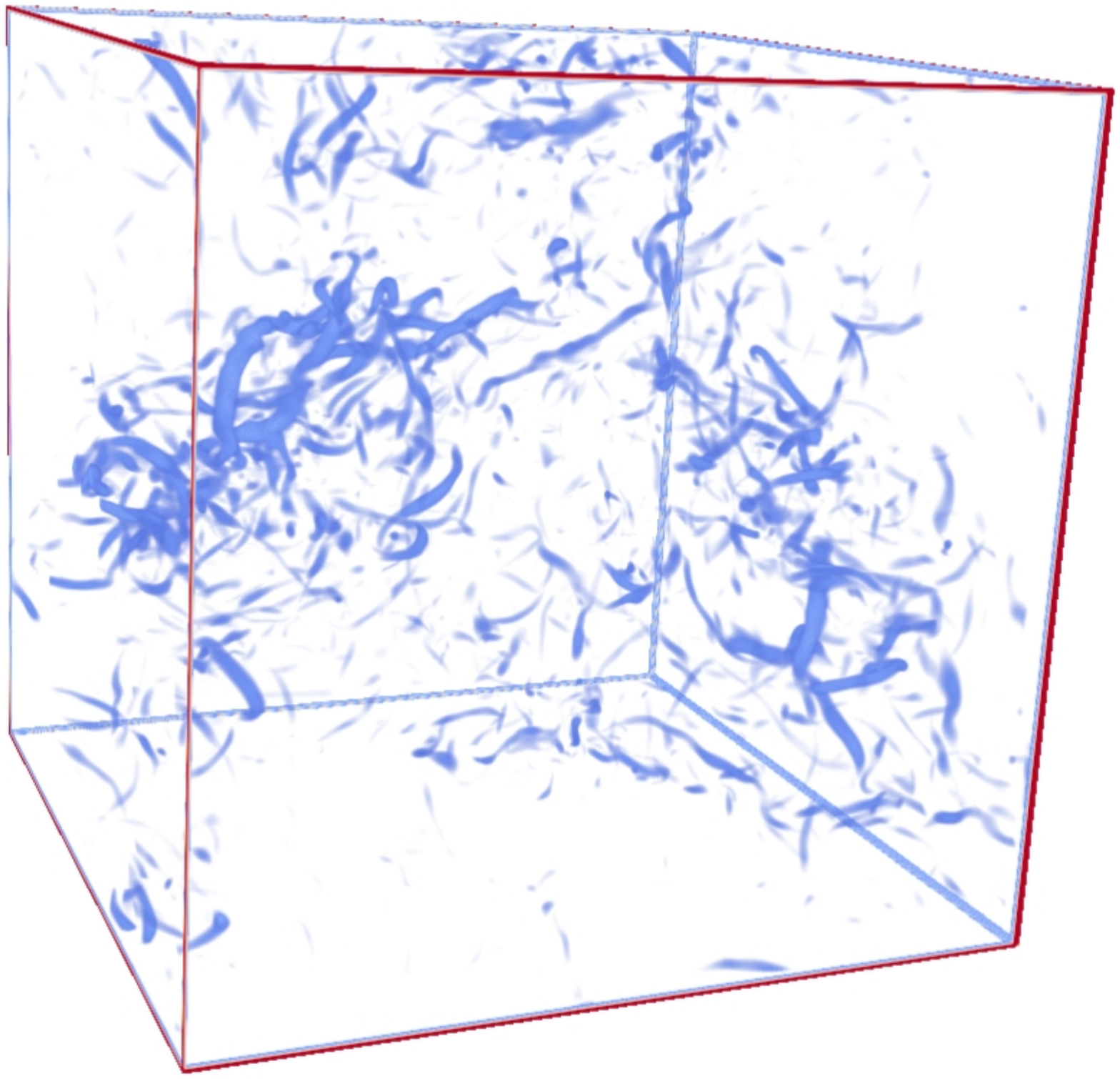}
\includegraphics[width=0.49\textwidth]{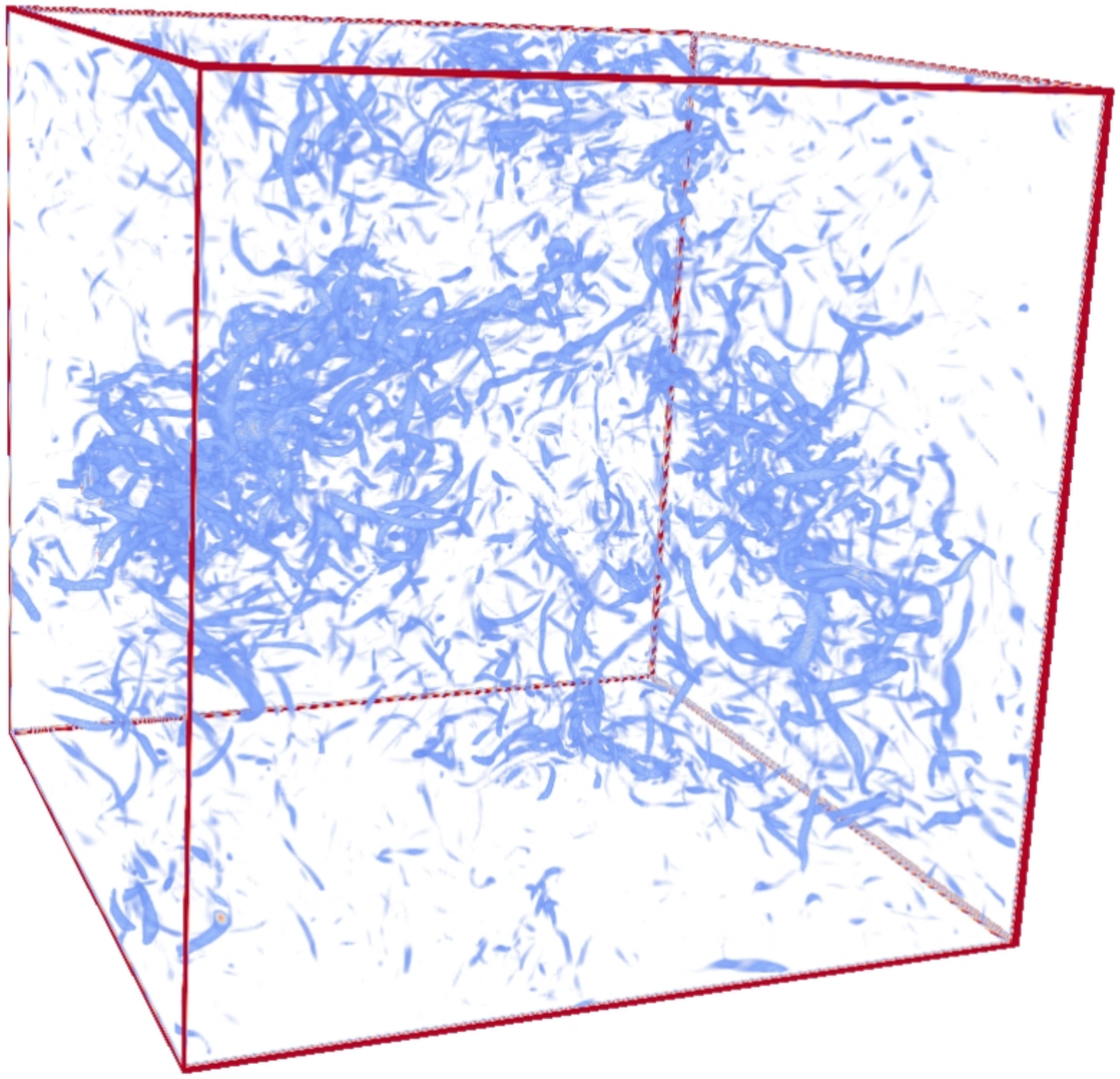}
\includegraphics[width=0.49\textwidth]{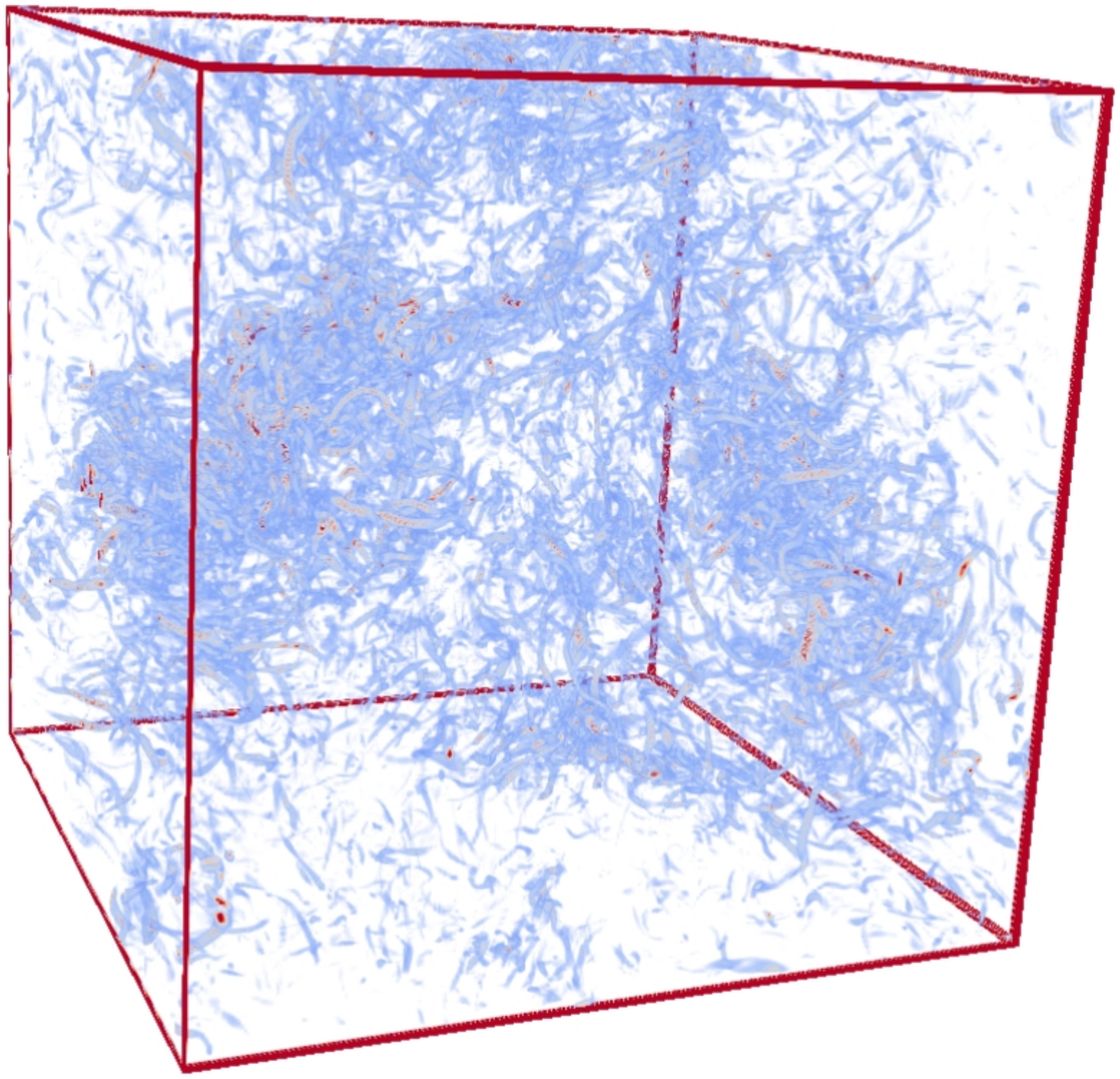}
\end{center}
\caption{The enstrophy of the filtered field for (a) $q=16, (b) q=32, (c) q=64$, using the Gaussian filter
and (d) unfiltered. \label{fig2}}
\end{figure*}
We consider now the qualitative phenomenology of the flow as represented through filtering at different scales. In Fig. \ref{fig2}, we show the vorticity field at different coarse-graining levels using the Gaussian filter. The filtering at different scales reveals a hierarchy of vortexes of different scales coexisting in the flow. The unfiltered field shows small vortex filaments typical of isotropic turbulence.  When filtered at small scales $q=64$ and also at inertial scale, the field shows some qualitative self-similarity even though smallest filaments are smoothed out. As the filtering wavenumber is reduced, larger and large vortexes are revealed. Even, when most of the scales are filtered out $q=16$, some residual elongated vortex tubes persist, pointing out the most important spatial, yet temporally intermittent, coherent structures of the flow. 


\subsection{Pdf of local fluxes} %
\begin{figure*}
\begin{center}
\includegraphics[width=0.49\textwidth]{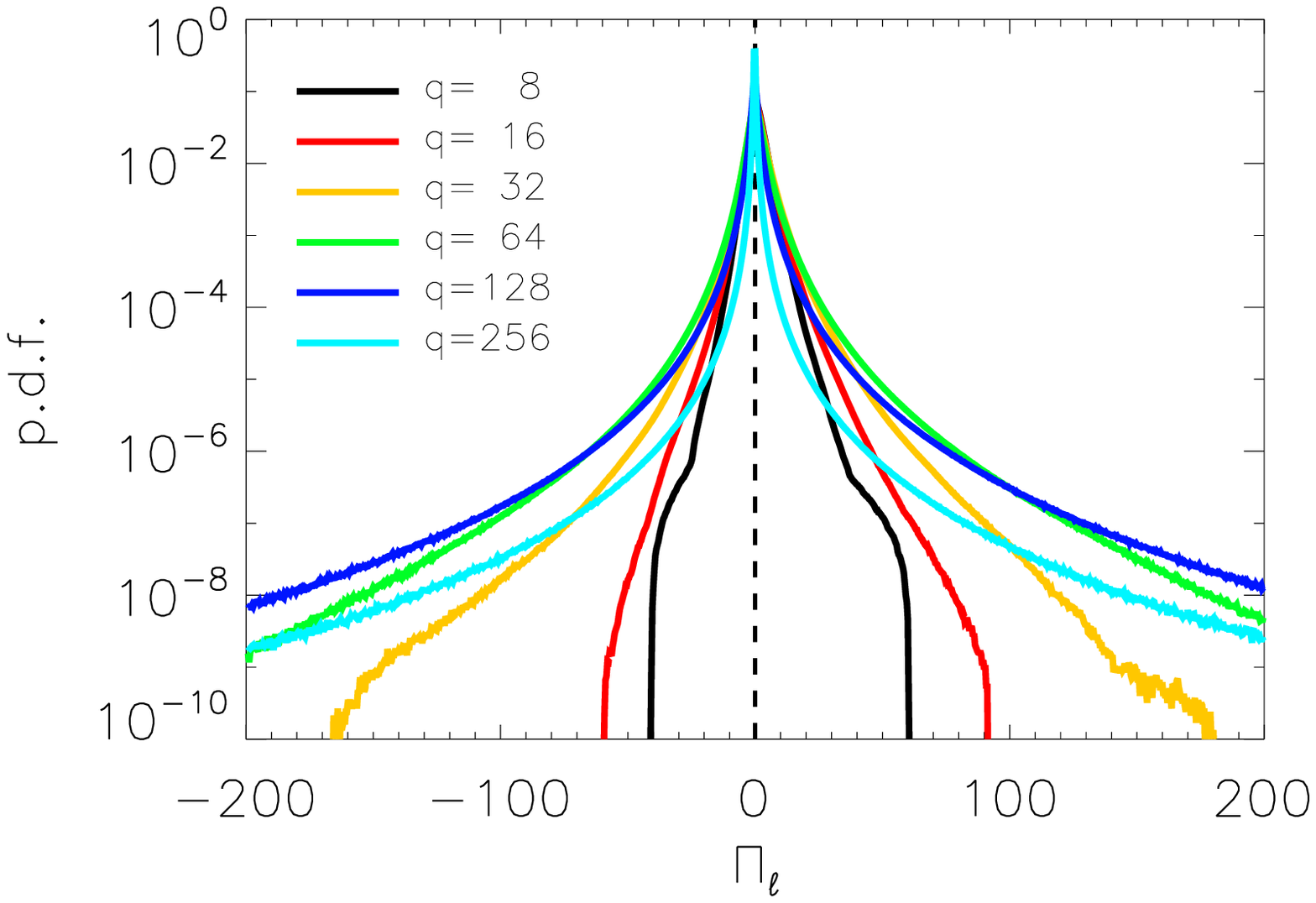}
\includegraphics[width=0.49\textwidth]{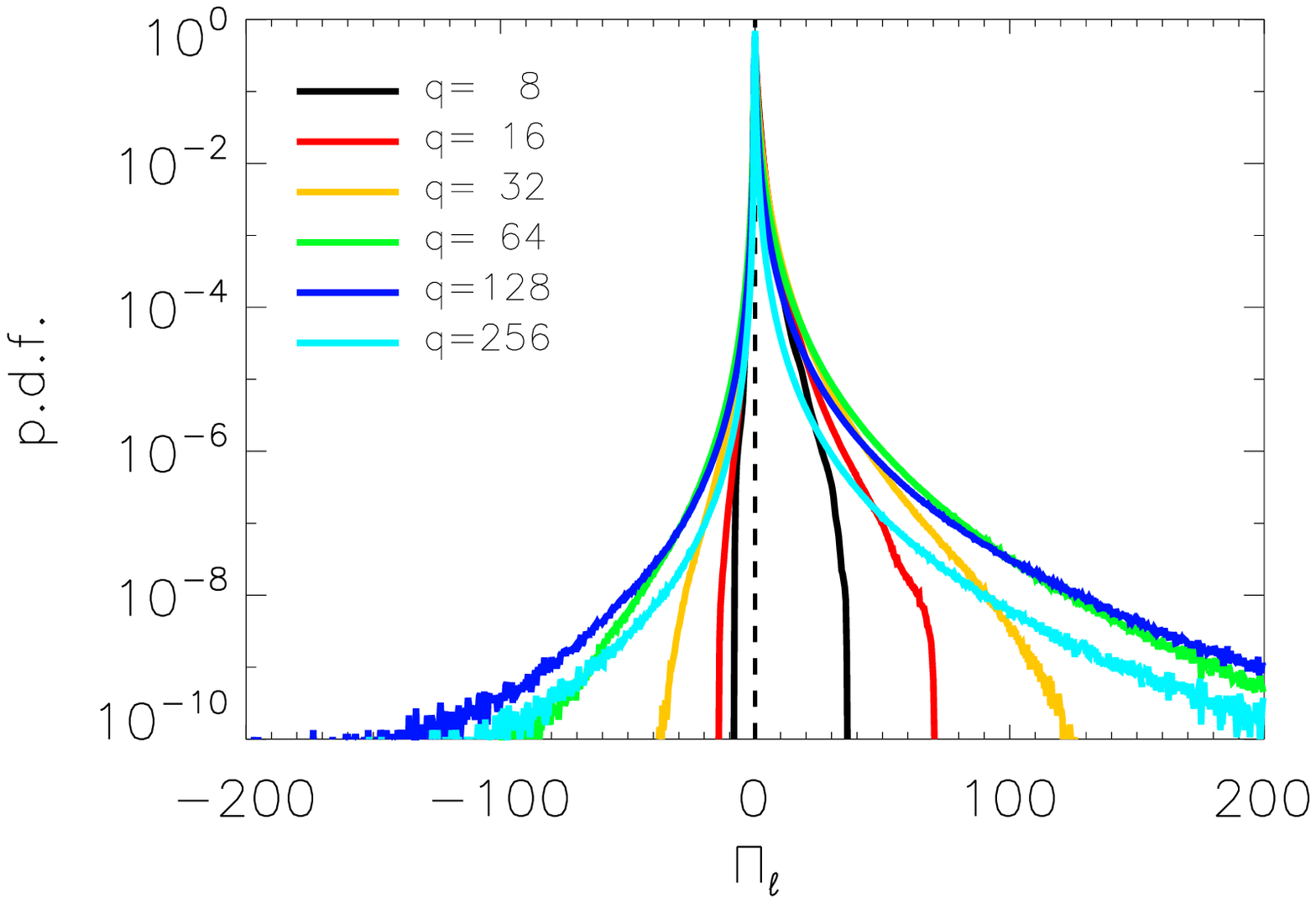}
\end{center}
\caption{The pdfs of the subscale energy flux for the sharp filter (left) and for the Gaussian filter (right). The pdfs are displayed for different length-scale cutoff $\ell$. }
\label{fig3}
\end{figure*}
\begin{figure*}
\begin{center}
\includegraphics[width=0.49\textwidth]{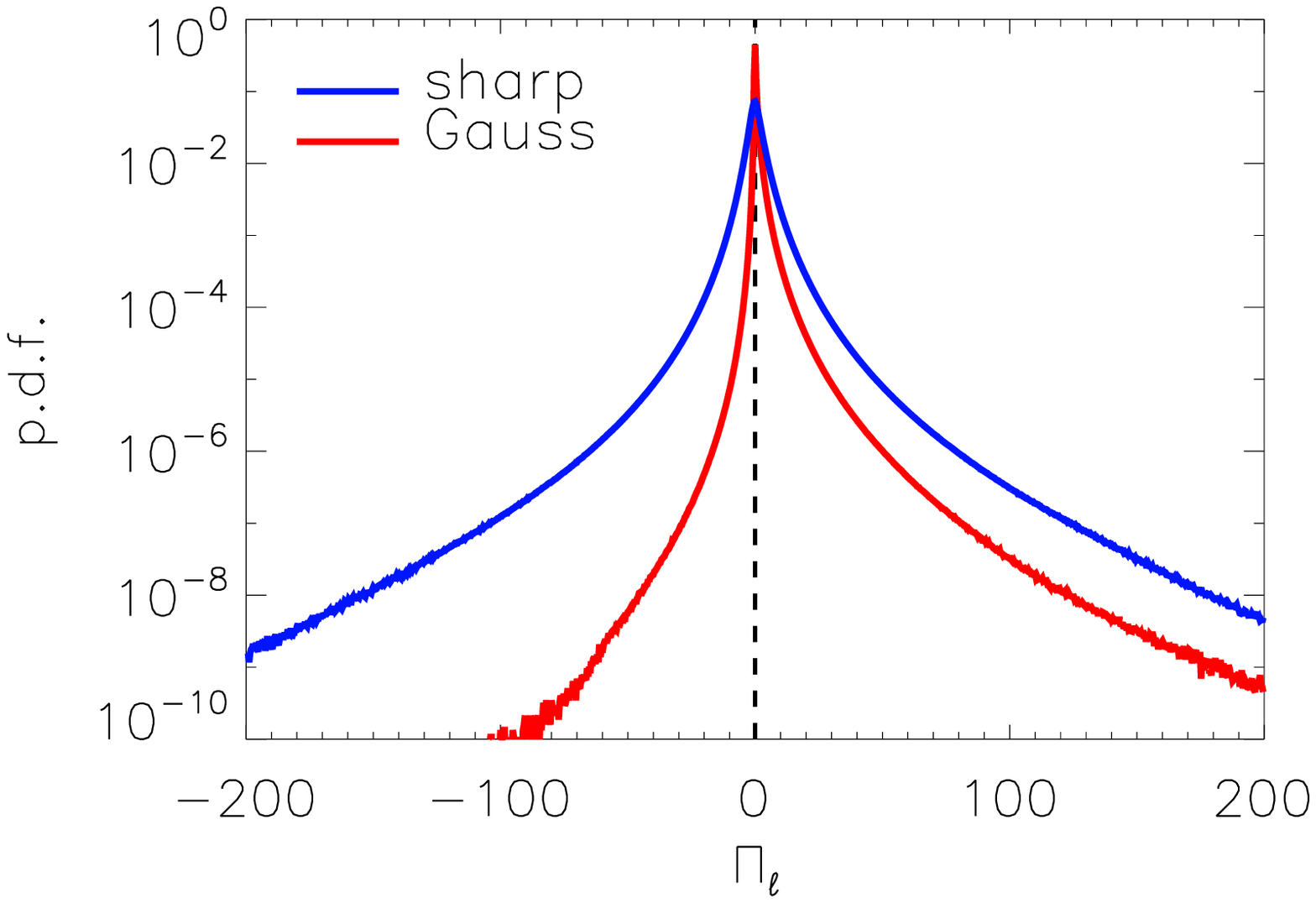}
\includegraphics[width=0.49\textwidth]{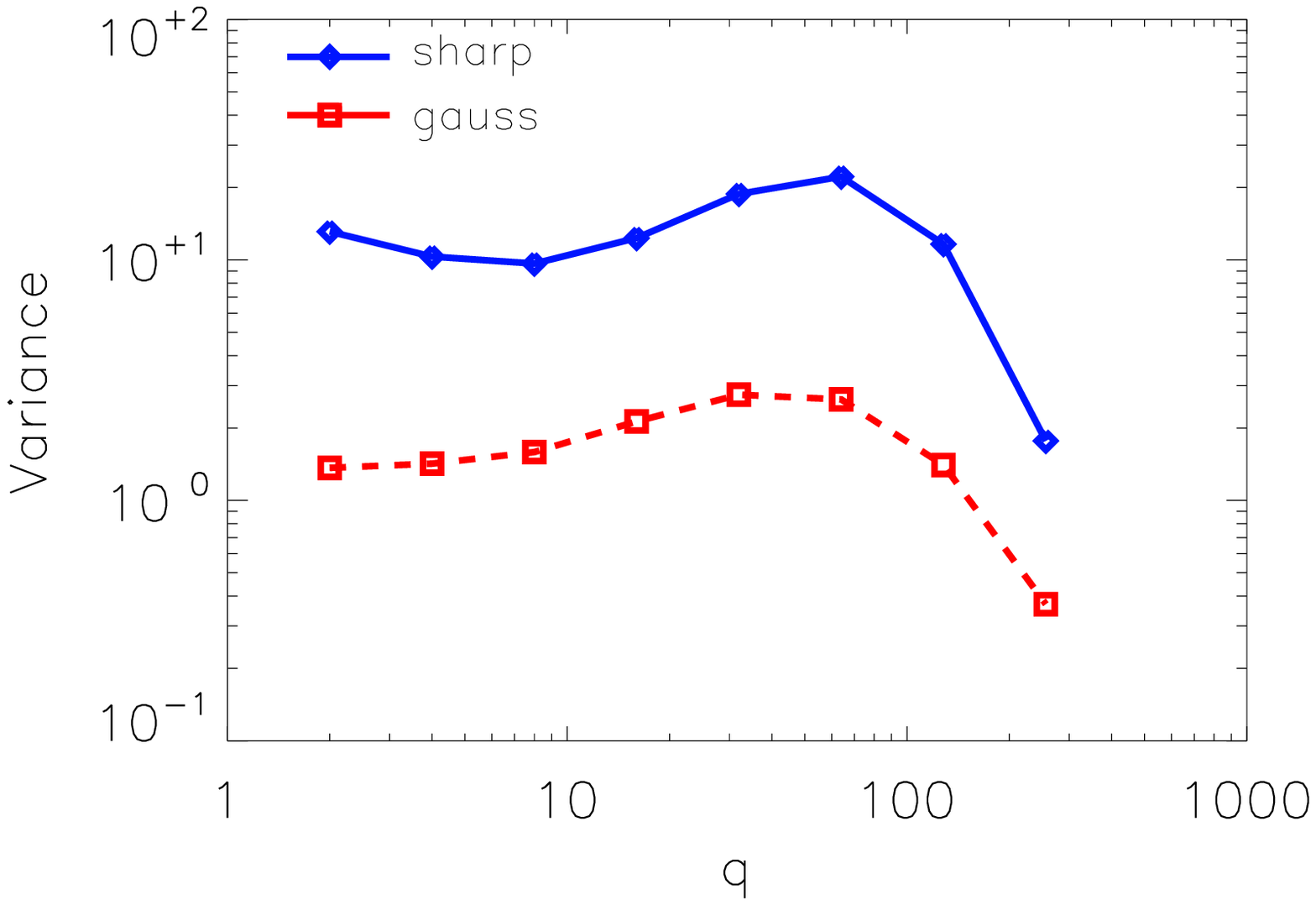}
\includegraphics[width=0.49\textwidth]{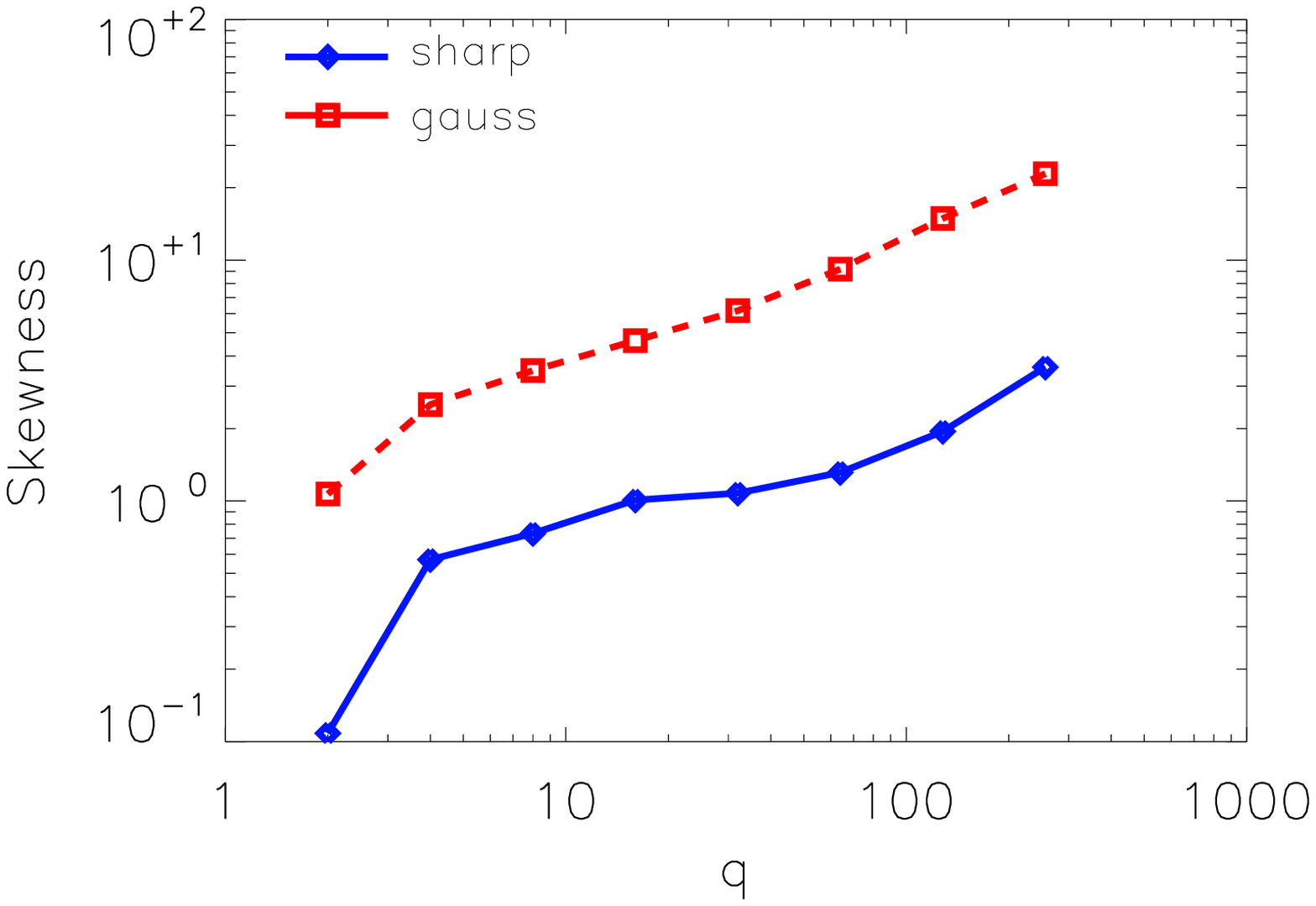}
\includegraphics[width=0.49\textwidth]{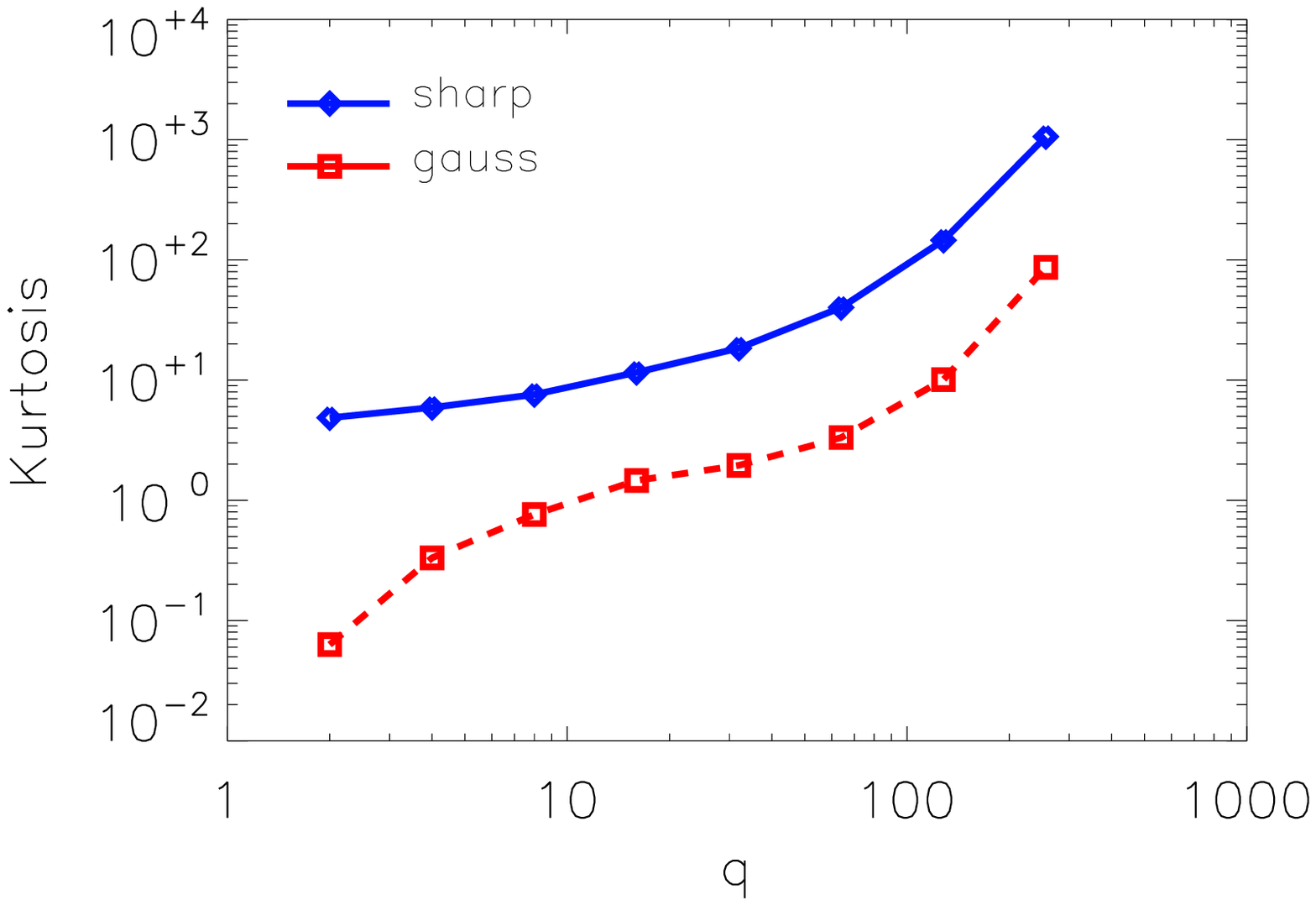}
\end{center}
\caption{Top left panel:A comparison of the pdf of the flux using the sharp filter (blue line) and a Gaussian filter (red line) for $q=64$. Top right panel: The variance of the flux as a function of $q$ for the two filters.
Bottom panels: The skewness (left) and the flatness (right) of the flux for the two filters.}
\label{fig4}
\end{figure*}

Using the flow described in the previous section we calculate the  local energy flux $\Pi_\ell({\bf x})$ of eq. (\ref{eq:filter}) at different levels of filtering $q=1/\ell$  using both the Gaussian and the sharp filter and analyse its statistical properties. The local flux was calculated for $q$ equal to powers of 2, $q=2^n$, where $n$ ranges from $1$ to $8$ as indicated in figure \ref{fig1}. This calculation was repeated for several instances of time so that
we obtained a good statistical sample. 
In Fig. \ref{fig3}, we show the pdfs of $\Pi_\ell$ at different scales, from the large ($q=8$) to the small ($q=256$) scales. 
For both filters the pdfs are centered around a value close to zero with long exponential or stretched exponential tails. The tails of the pdf increase as $q$ is increased up until the dissipation scales are reached
$q\simeq 128$ after which they start to decrease. It is worth noting  that for $q > k_{max}$ the local flux is point-wise zero so the pdf converges to a delta function at $\Pi_\ell(x)=0$.

The flux obtained by sharp filtering is more symmetric and displays larger tails. The flux obtained through Gaussian filtering has a more skewed behaviour, with less probable negative events. The profiles obtained are similar to those obtained in analogous previous simulations \cite{chen2003joint,buzzicotti2018effect,domaradzki2007comparison}. 
In the top right panel of Fig. \ref{fig4} we compare the two fluxes based on the two filters at the scale $\ell=1/q=1/64$, which is at the end of the inertial range and emphasises the differences of the pdfs of the two filters. In the rest of panels of fig. \ref{fig4} we show the first normalized statistical moments of the flux:
\be 
\mathrm{the\,\,variance}\quad
\langle \left[\Pi_\ell - \langle \Pi_\ell  \rangle\right]^2 \rangle, \ee 
\be 
\mathrm{the\,\,Skewness}\quad
\frac{
 \langle \left[\Pi_\ell - \langle \Pi_\ell  \rangle\right]^3 \rangle}
{\langle\left[ \Pi_\ell - \langle \Pi_\ell  \rangle\right]^2 \rangle^{3/2}}
\qquad
\mathrm{and\,\, the\,\,Kurtosis}\quad
\frac{
 \langle\left[ \Pi_\ell - \langle \Pi_\ell  \rangle\right]^4 \rangle}
{\langle \left[\Pi_\ell - \langle \Pi_\ell  \rangle\right]^2 \rangle^{2}}, 
\ee
computed at different scales and for the two filters. 
We have here a vivid description of the difference between the two results. The flux computed through the Gaussian filter at this scale is characterised by strong fluctuations with possible but rare negative events.
It is interesting to point out that the shape exhibited by the one-point pdf of the flux shown in Fig. \ref{fig4} is qualitatively the same to those found in the study of general dissipative non-equilibrium systems, in the framework of the Gallavotti-Cohen or fluctuation-relation analysis \cite{evans1993probability,gallavotti1995dynamical,Aumaitre:2001p6195,ciliberto1998experimental,shang2005test,falcon2008fluctuations,marconi2008fluctuation,bandi2009probability,zonta2016entropy}.
Instead, the negative events (or backscatter events, as called in LES) are much more frequent with the sharp filter. It is interesting to look at the statistical moments, which show that the two approaches give the same trend up to the fourth moment at all scales, within the numerical errors, but there is about one order of magnitude of difference between the two results almost everywhere. 
While the average flux is the same computed by the two different methods, the sharp filter gives a wildly fluctuating subscale energy flux, with many negative events, so that it appears difficult to use directly in the framework of LES, at least from a numerical point of view. The reason of such discrepancy is traced back to the fact that the sharp filter used here has not the desirable features of a proper filter, notably is not localised in space and is not positive-definite.
Even though it has been shown that the sharp spectral filter has a firm theoretical basis for its use in LES,
 the results suggest that the use of the smooth filtering approach is preferable if interested in energy flux properties.
For this reason, we focus in the following on the Gaussian filter.

\subsection{Joint pdf} 
\begin{figure*}
\begin{center}
\includegraphics[width=0.49\textwidth]{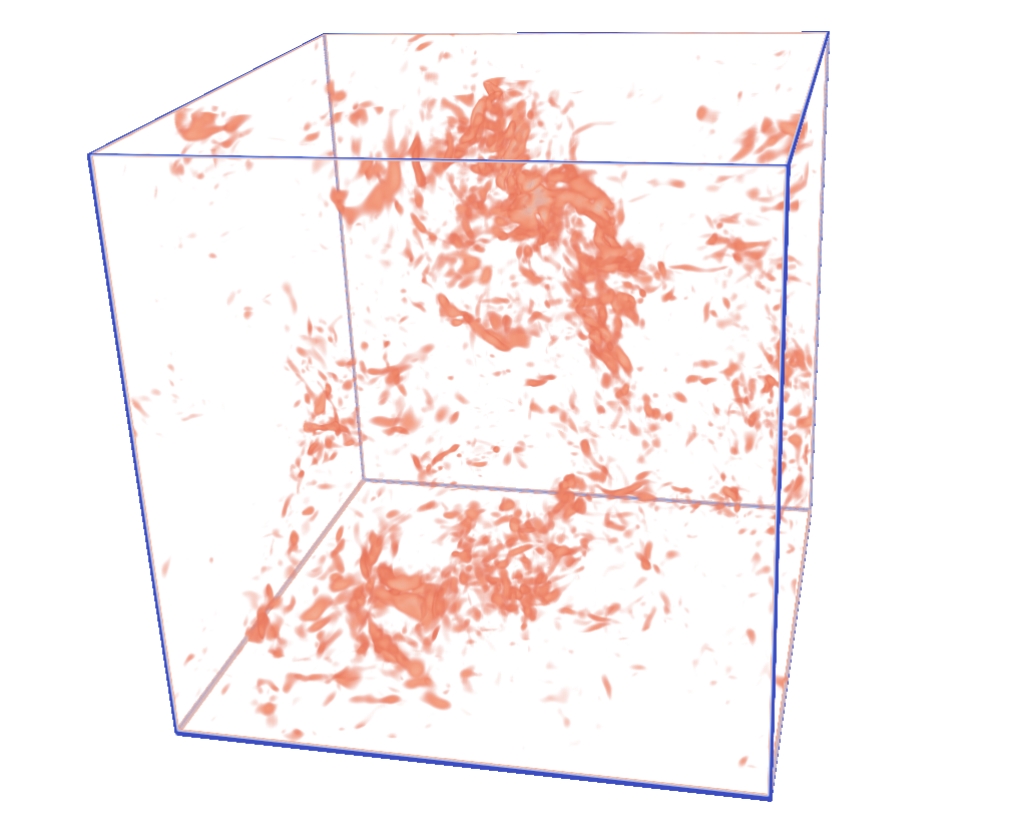}
\includegraphics[width=0.49\textwidth]{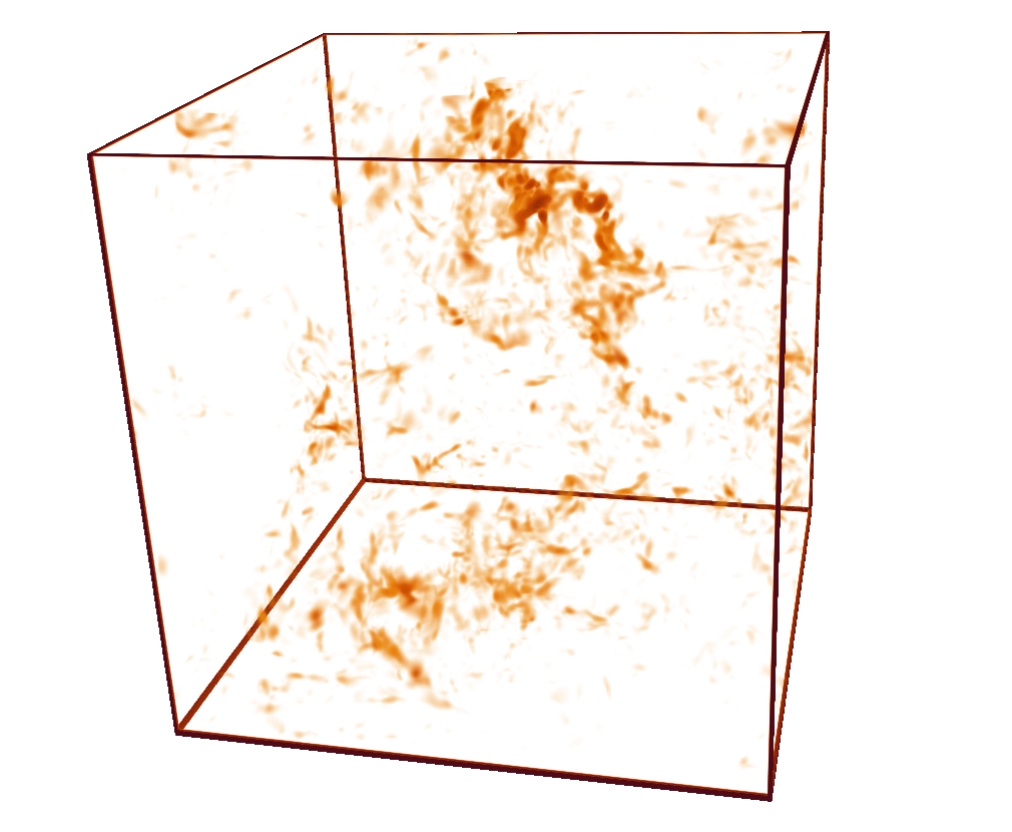}
\end{center}
\caption{ A visualisation of the 
strain density (left) and the amplitude of the local flux (right), for$ q=32$ and Gauss filtering (for the same snapshot of the flow as is fig \ref{fig2}). }
\label{fig5}
\end{figure*}
\begin{figure*}
\begin{center}
\includegraphics[width=0.40\textwidth]{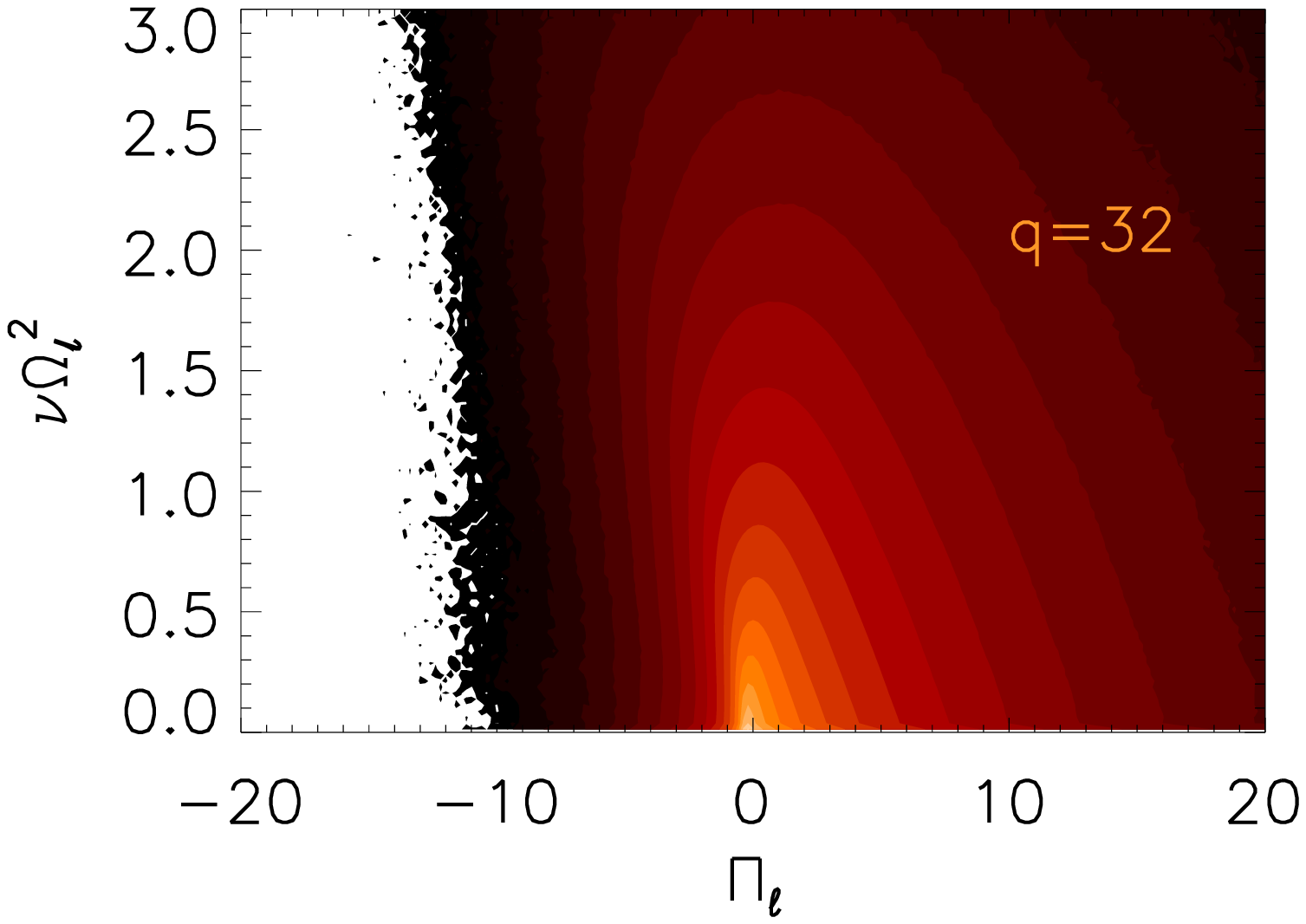}
\includegraphics[width=0.40\textwidth]{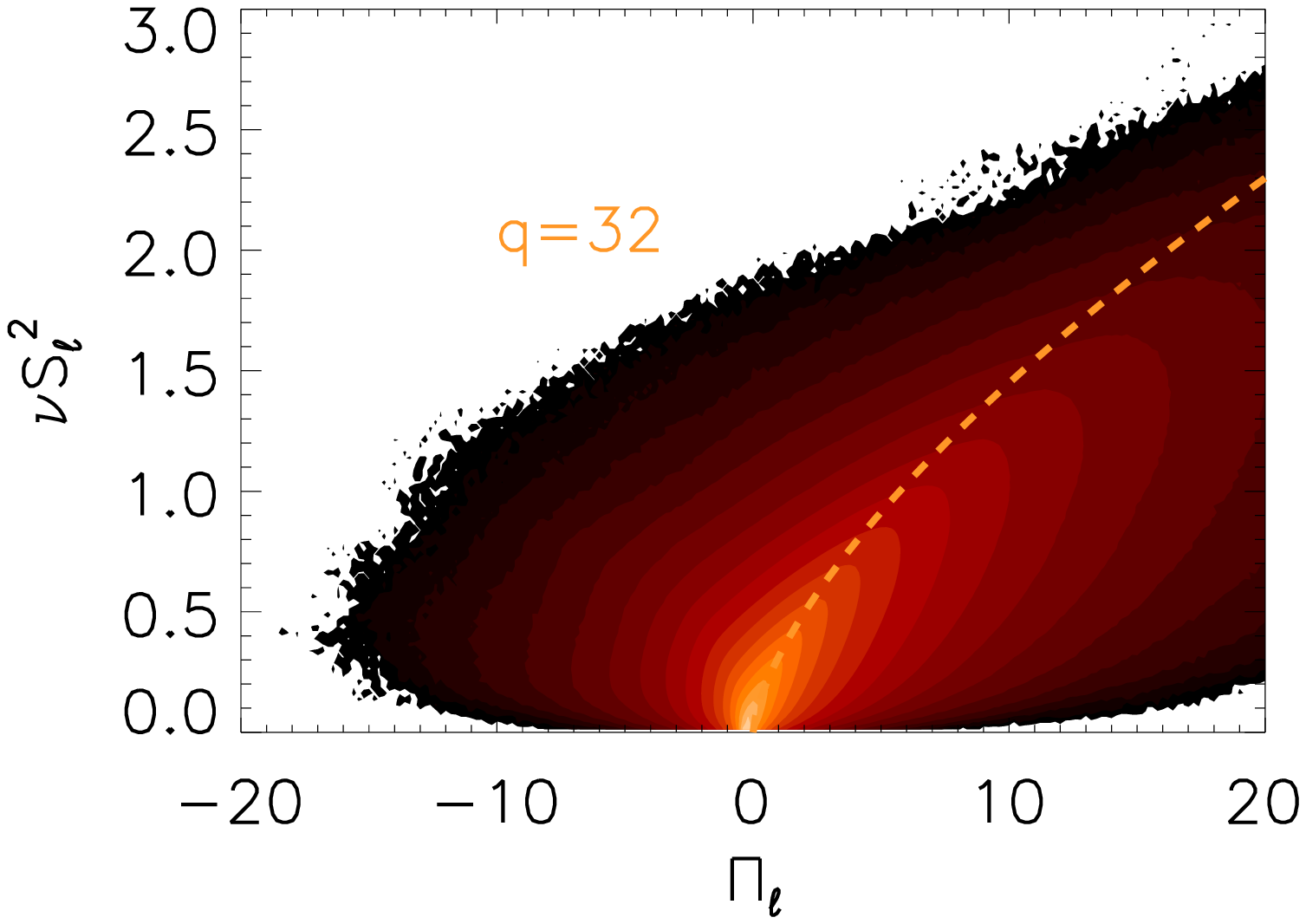}\\
\includegraphics[width=0.40\textwidth]{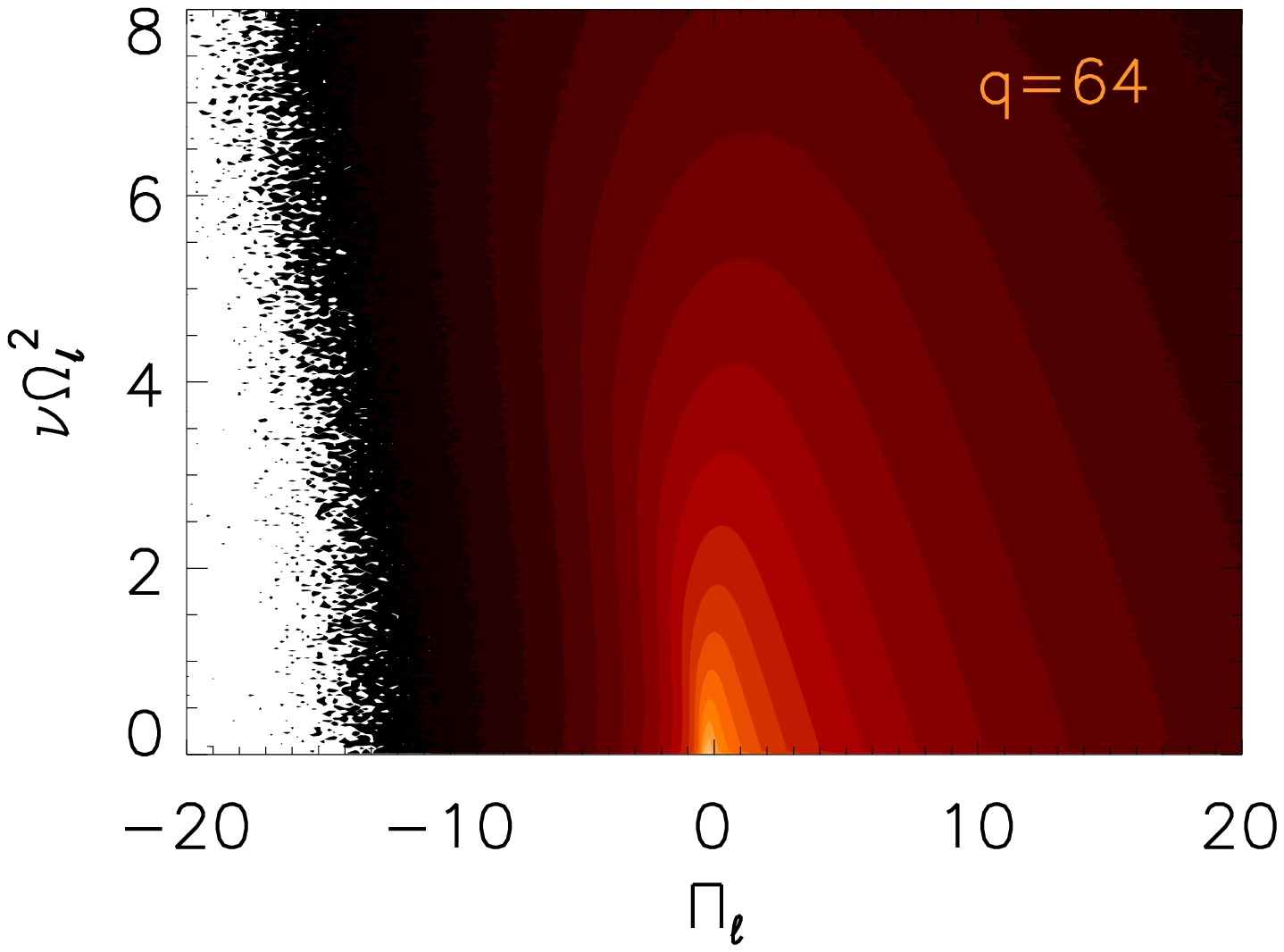}
\includegraphics[width=0.40\textwidth]{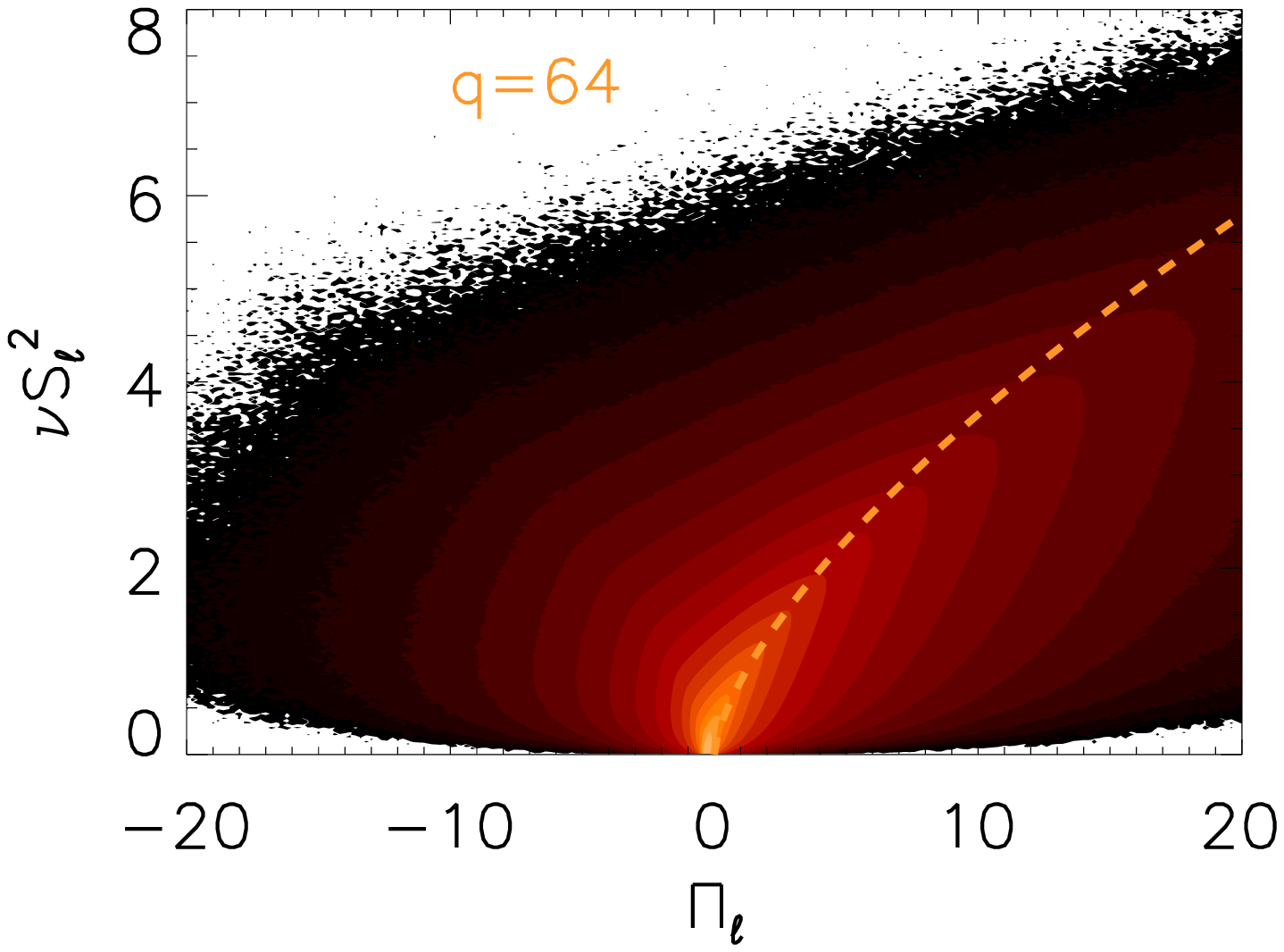}\\
\includegraphics[width=0.40\textwidth]{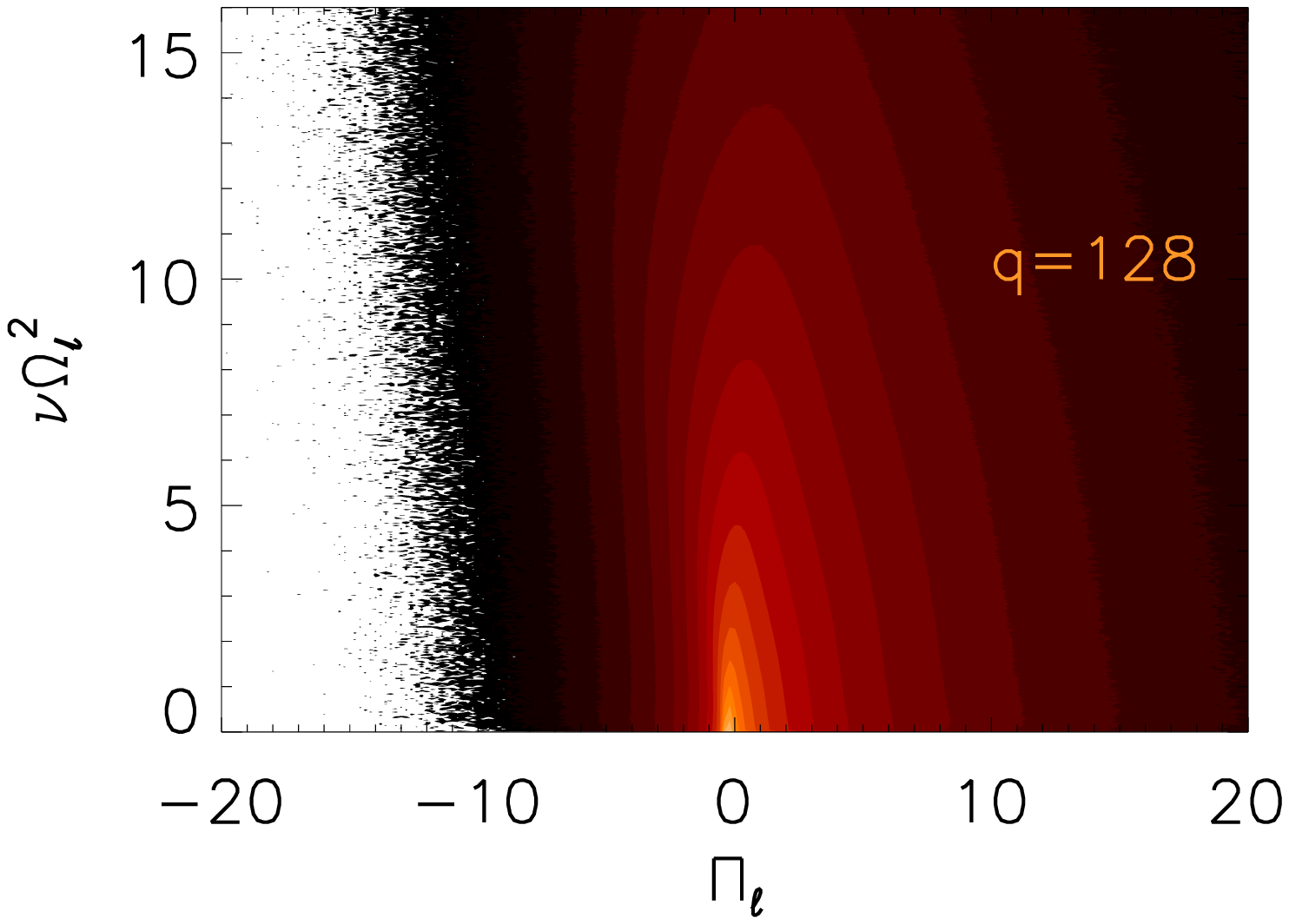}
\includegraphics[width=0.40\textwidth]{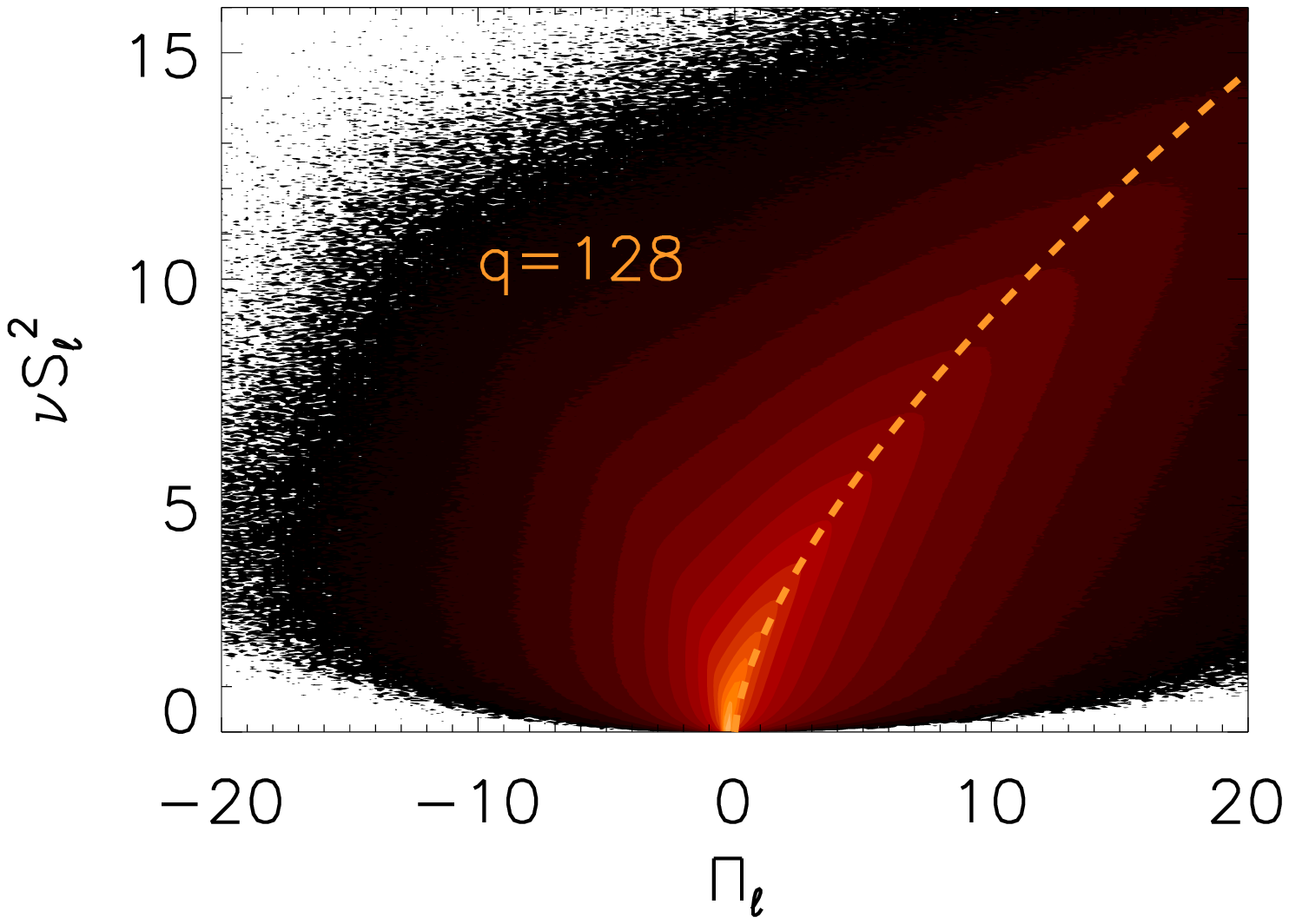}\\
\includegraphics[width=0.40\textwidth]{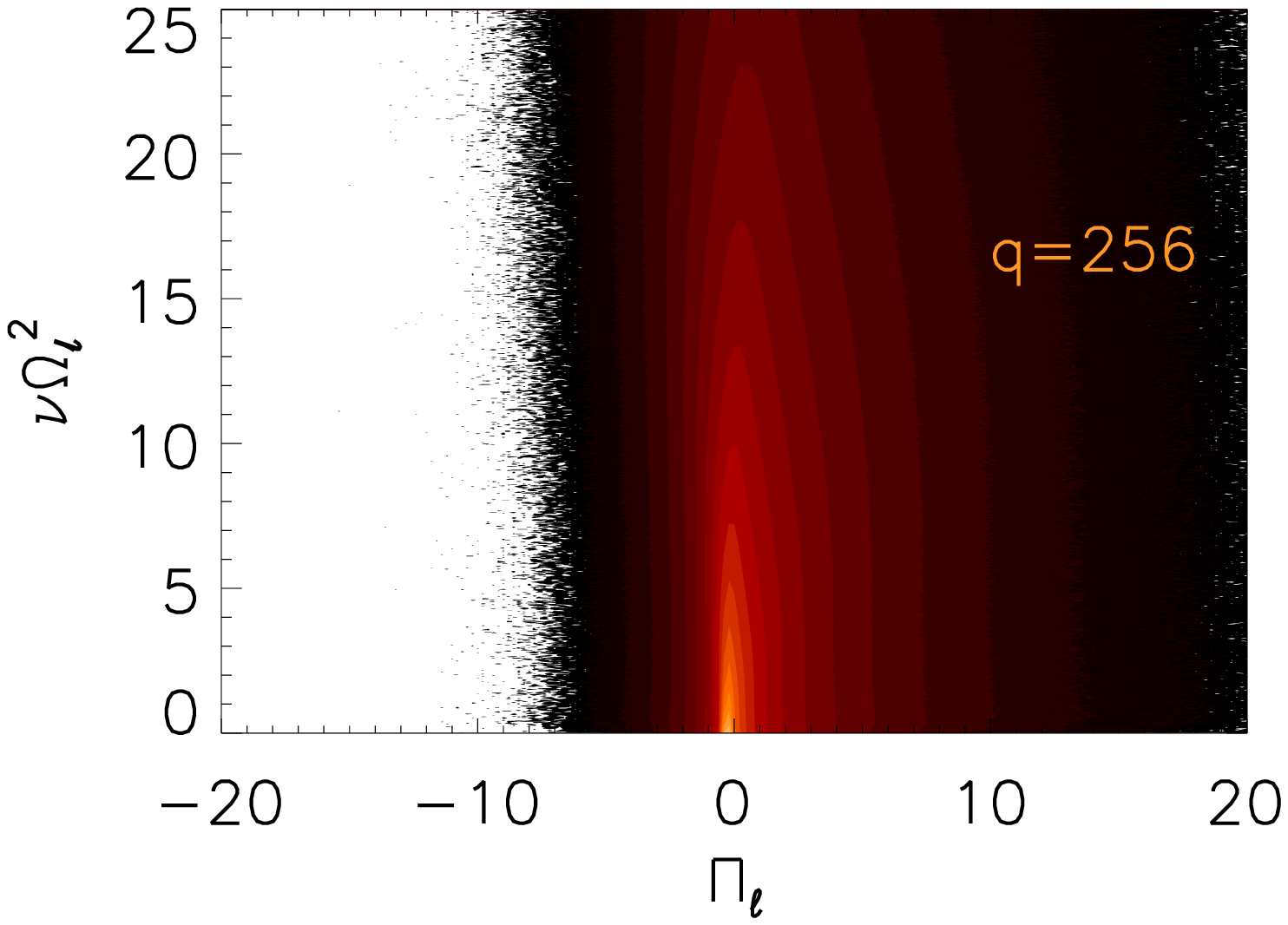}
\includegraphics[width=0.40\textwidth]{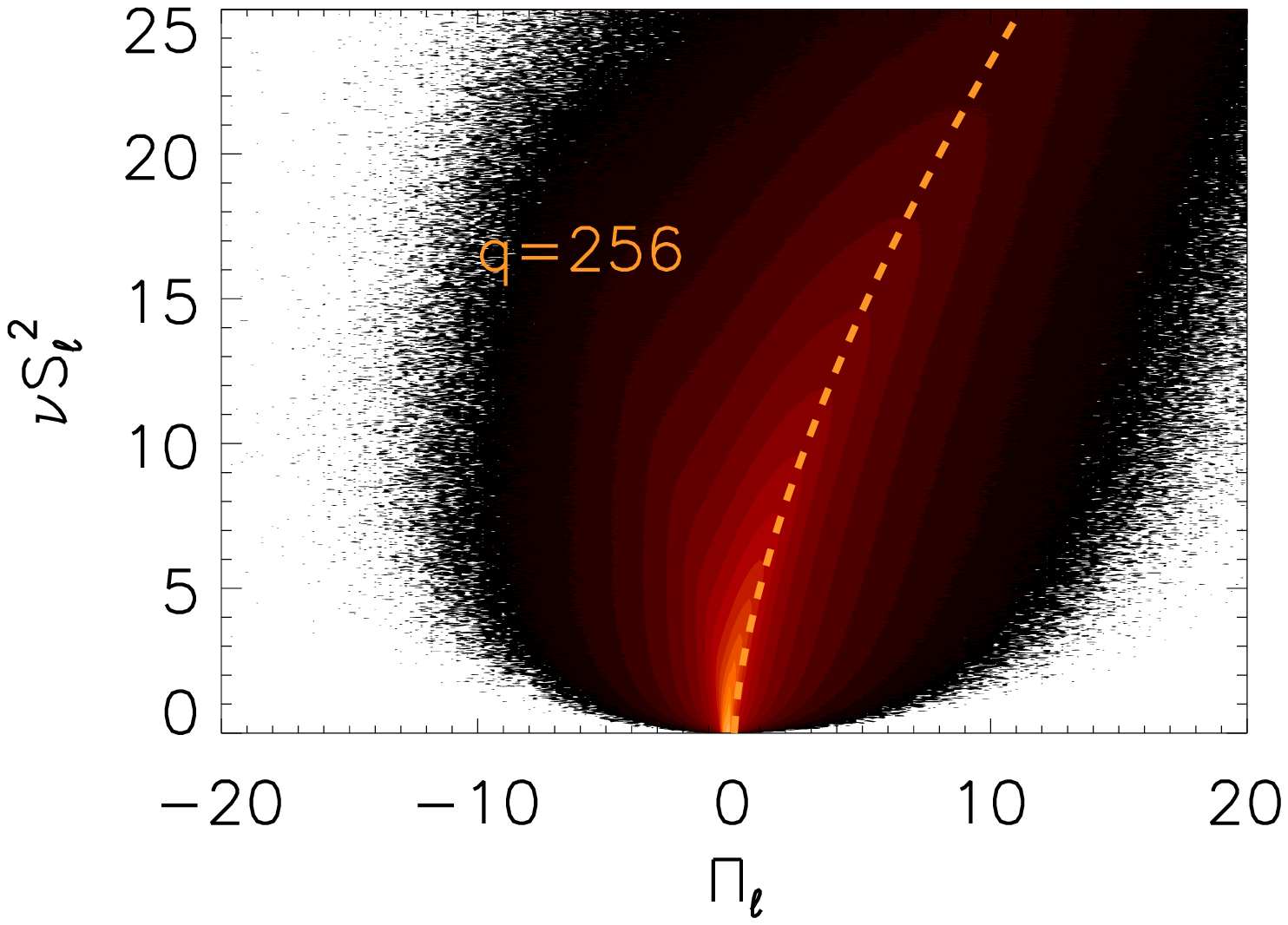}
\end{center}
\caption{
Left column: joint pdf of flux $\Pi_\ell$ and enstrophy $\Omega^2$ at different scales. Right column: joint pdf of flux $\Pi_\ell$ and strain $S^2$ at different scales. Bright colors indicate high probability while dark colors indicate low probability, white indicates zero probability. The strain and and the enstrophy are multiplied with the viscosity to be able to compare with the mean value at $\ell=0$ given by $\nu \lra{S^2}=\nu \lra{\Omega^2} =\epsilon=1$.}
\label{fig6}
\end{figure*}


We now look at the possible dependence of the local energy flux with the gradients of the flow as discussed in section \ref{modeling}. To give some more visual insights, in figure \ref{fig5} we show a three dimensional visualisation of the coarse-grained strain $S_\ell$ and the energy flux $\Pi_\ell$ at the same scale side-by-side. We have chosen a moderately large scale $q=32$, since at these scales the dependence on vorticity seems more important. The figure convincingly shows that much of the properties of the flux and notably the geometrical features are well reproduced by the strain. In particular $\Pi_\ell$ appears to be more correlated with the filtered strain than the filtered enstrophy shown in the top right panel of fig. \ref{fig2} for the same $q$.  

To be more quantitative we calculate the joint pdf between $\Pi_\ell$ and $\Omega_\ell^2$ and  between $\Pi_\ell$ and $S_\ell^2$. The strain and and the enstrophy are multiplied with the viscosity to be able to compare with the mean value at $\ell=0$ given by 
\[ \lim_{\ell\to0}\nu \lra{S_\ell^2}      = 
   \lim_{\ell\to0}\nu \lra{\Omega_\ell^2} =
   \epsilon=1 .\]
The results are displayed in figure \ref{fig6}. 
From the left column, we can see that the energy flux is essentially uncorrelated with the enstrophy, and therefore with vorticity. In particular for large $q$ we observe that at 
a given $\Pi_\ell$ the probability changes very slowly with respect to $\Omega$, indicating almost independence.
The joint pdf of the flux with the strain present a very different story. The two variables appear very strongly correlated at each scale. Furthermore, even though the change of scale has an impact on the shape of the pdf, it appears to change in a self-similar way, that is the dependence on the scale is given by a power law.
To quantitatively capture this trend, we show in all figures the curve given by the Smagorinsky model (\ref{Smago}).
The agreement of this curve with the maximum of the probability is excellent. Qualitatively, regions with high strain favour large energy flux. Furthermore, as expected the larger the scales, the less important the strain can be, so that very large strain values are obtained at very small scales, where they contribute to the viscous dissipation. It is worth noting that the variations around the maximum value (given approximately by eq (\ref{Smago}) are significant and $\Pi_\ell$ also takes negative values. This is most significant for small values of $S$ while for large values of $S$ the flux is almost always positive.

\begin{figure*}
\begin{center}
\includegraphics[width=0.48\textwidth]{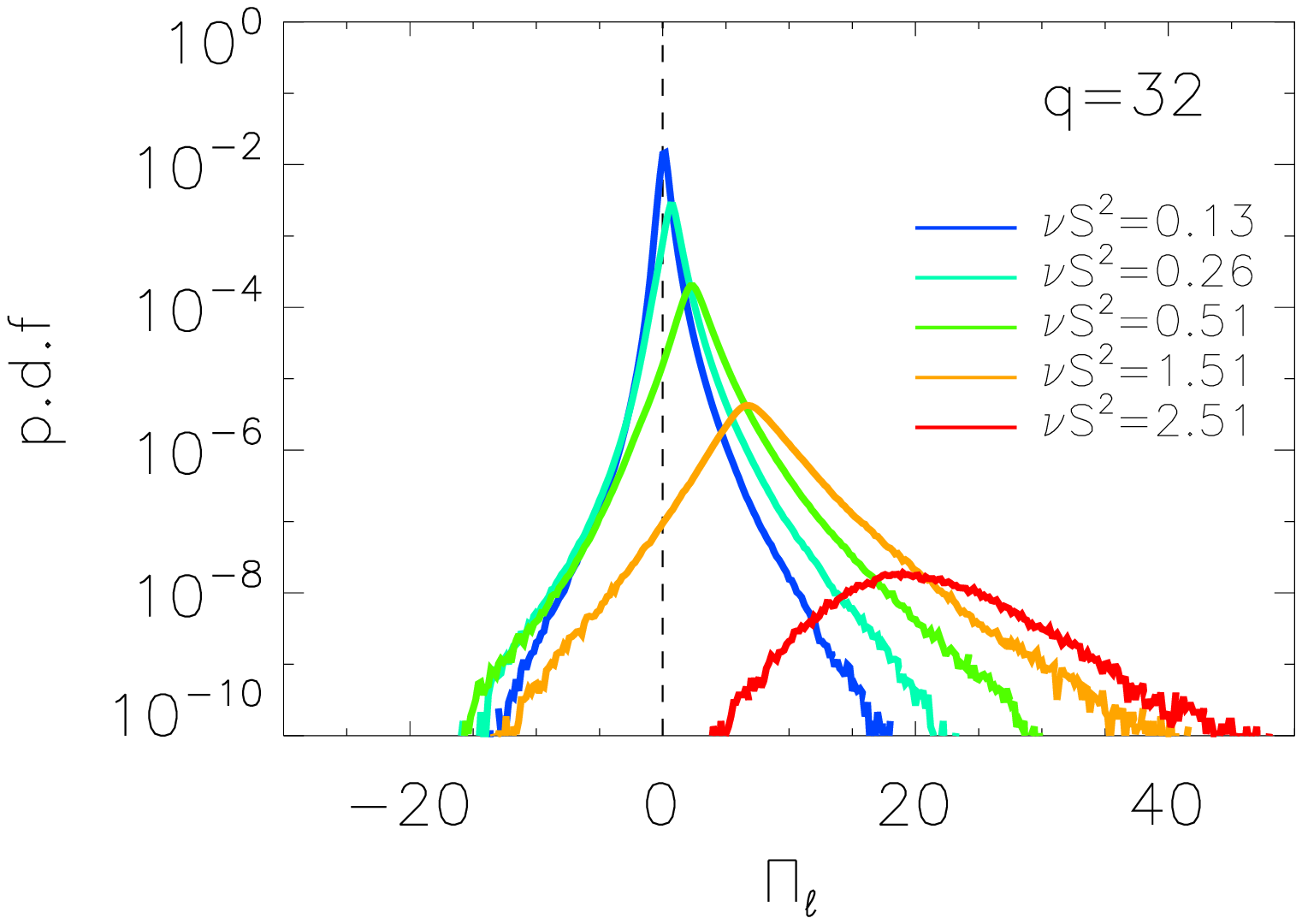}
\includegraphics[width=0.48\textwidth]{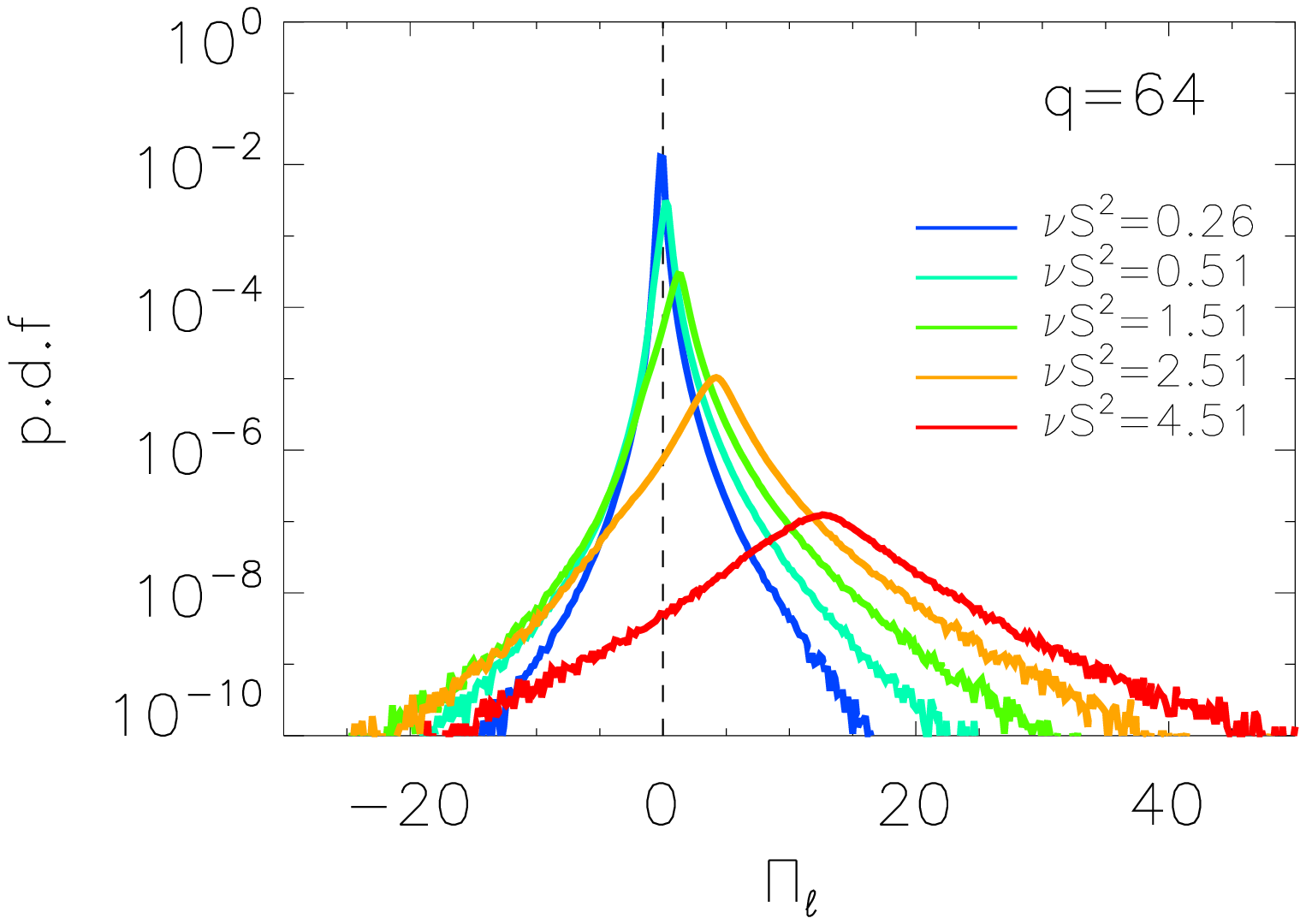}
\end{center}
\caption{ The pdf of the flux conditioned on different values of the strain $S_\ell$ for $q =32$ (left) and  $q=64$ (right). }
\label{fig7}
\end{figure*}
To get more insights on the behaviour of the fluctuations, we plot the shape of the pdf of the sub-scale flux $\Pi_\ell$ at a given scale conditioned with several values of the strain,  shown in Fig. \ref{fig7} for the scale $\ell=1/q=1/32$ and $\ell=1/64$. While the mean value and the maximum follow the $S_\ell^{3}$ curve, the shape of the curves changes even at the qualitative level. Indeed, positive extreme events are found only for large strains, which means more pronounced right tails for the corresponding pdfs. 
Although the analysis focus on rare events and therefore statistical errors may induce to wrong conclusions, it is interesting to make the following remarks:
(i) the negative side of the flux is less affected by changes in $S_\ell^2$, at small values of the strain;
(ii) however, as larger values of $S_\ell^2$ are examined less negative events are observed, notably at the larger scale shown $q=32$.
Thus, large strain regions are related dominantly to positive flux;
(iii) At different scales, the fluctuations of the flux display a different behaviour, in particular when conditioned on high value of the strain.

\begin{figure*}
\begin{center}
\includegraphics[width=0.49\textwidth]{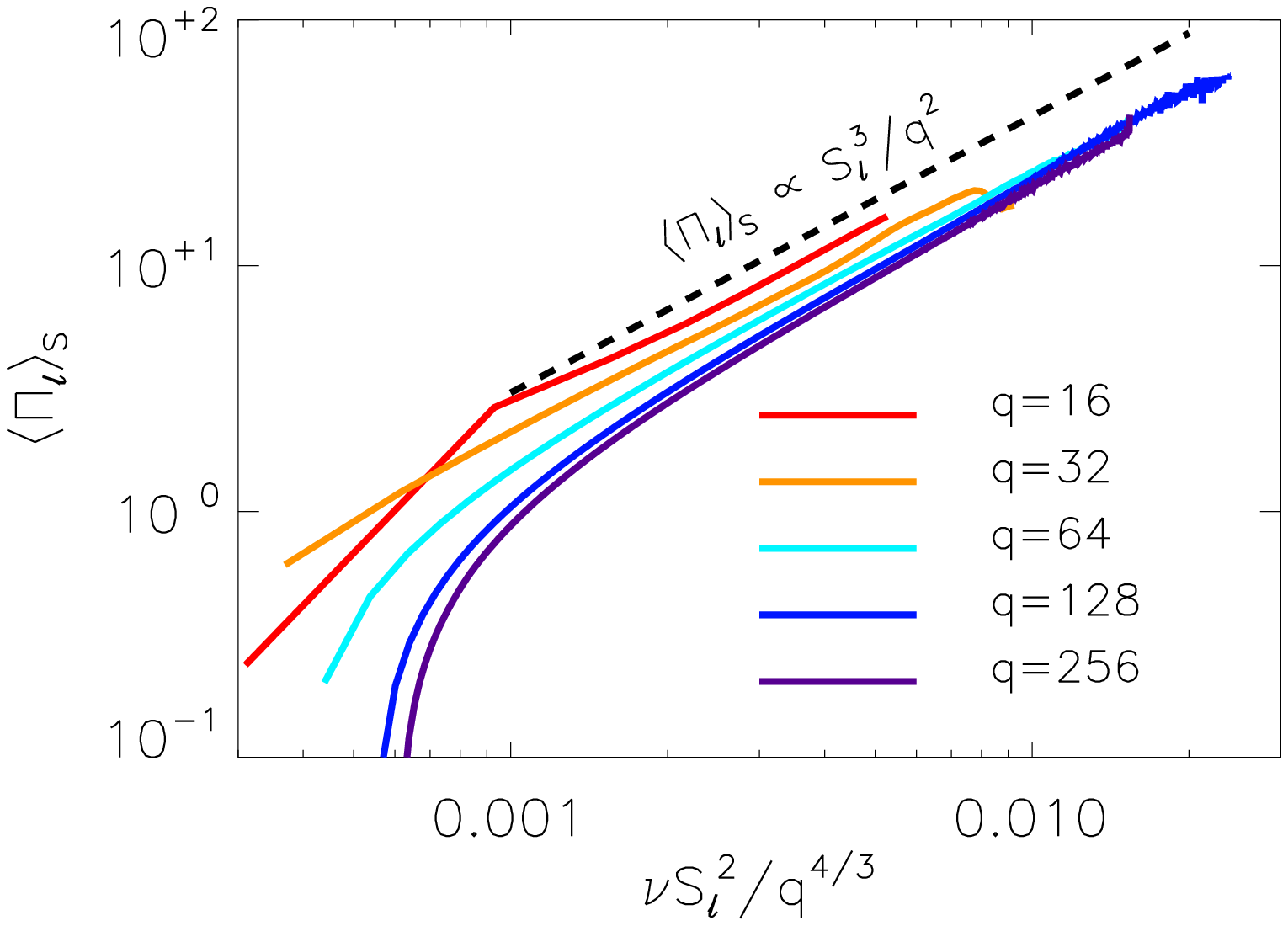}
\includegraphics[width=0.49\textwidth]{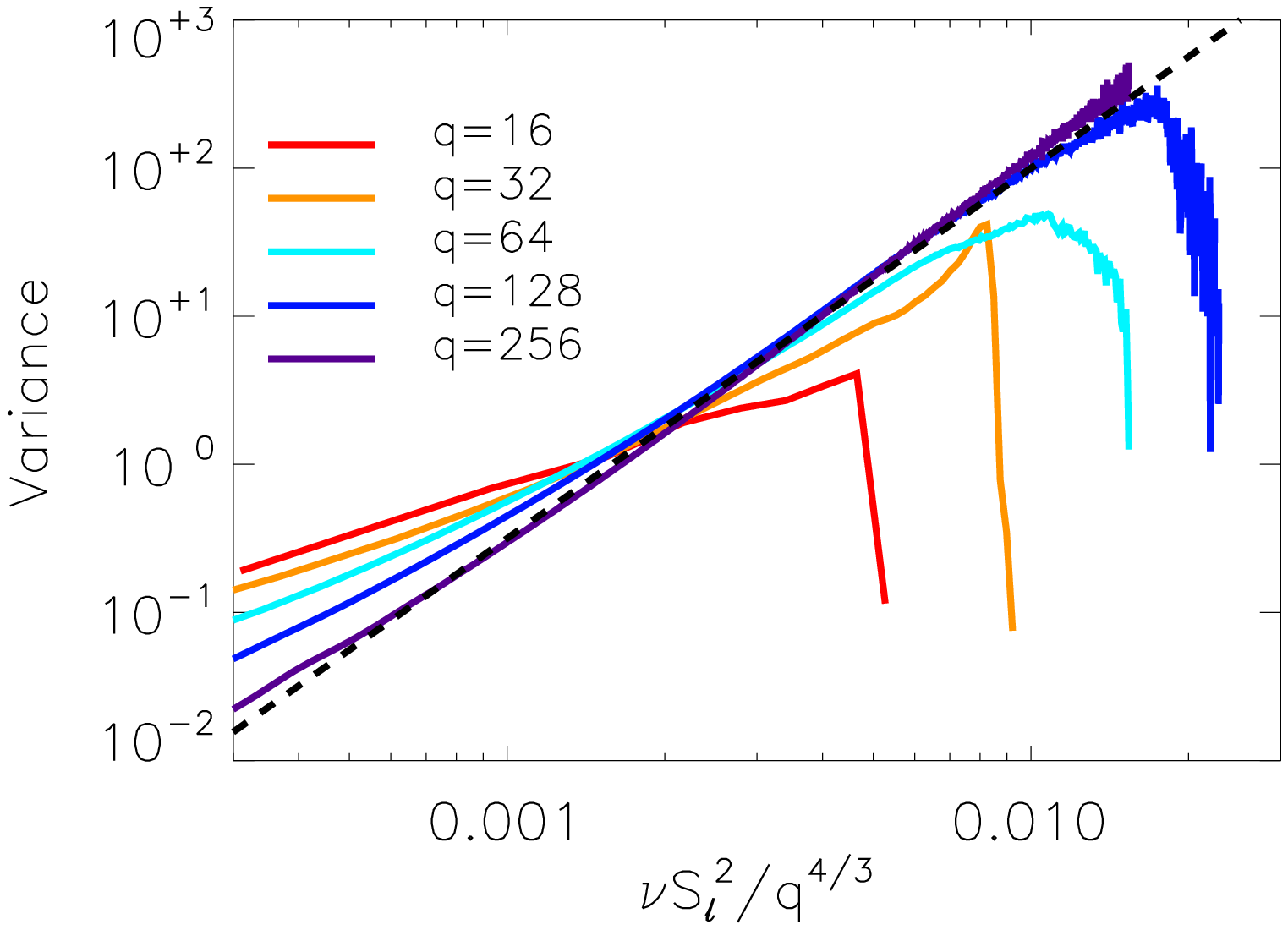}
\end{center}
\caption{ Left: the mean conditioned flux $\lra{\Pi_\ell}_S$ as a function $S_\ell$, Right: the variance of the conditioned flux as a function of $S_\ell$.}
\label{fig8}
\end{figure*}

Furthermore to examine the dependence on scales, as highlighted in Fig. \ref{fig6}, we have analysed the self-similarity of the behaviour and the results are displayed in figure \ref{fig8}. The left panel shows as a function of $\nu S_\ell^2$ the mean value of the flux conditioned on the strain $\langle \Pi_\ell\rangle_S$, where the average is performed over all points that have a given strain $S_\ell$. The curves nicely collapse, pointing out the behaviour indicated by eq. (\ref{Smago}). The results of the mean value thus suggest that the mean flux can be estimated and thus modeled by the value of the strain rate following the relation (\ref{Smago}). However, besides the mean value a successful model should also capture the  fluctuations around it. This is particularly important in this case since as shown in fig. \ref{fig7} although $\langle \Pi_\ell \rangle_S$ is always positive the fluctuations are strong enough that $\Pi_\ell$ also takes negative values.

To see if a similar relation is followed by the fluctuations and look for a possible self-similar behavior of the fluctuations we plot the variance of the conditioned $\Pi_\ell$ in the left panel of Fig. \ref{fig8}. The behaviour of the variance is not fully self-similar, as already pointed out by figure \ref{fig7}. 
As $q$ is varied different slopes are observed. For scales in the inertial range and larger $q\le 64$, the 
slope observed in the left panel of fig.\ref{fig8} (indicating a possible power-law dependence) is decreasing with $q$. Only in the dissipative range $q>64$, the curves almost collapse. This lack of self-similarity is a clear sign of intermittency that resists a theoretical understanding till today and requires further investigations.  

\section{Discussion}

The main focus of the present multi-scale analysis of the cascade energy process is on giving some insights in relation to LES of turbulent flows, where only scales larger than $\ell$ are simulated and their effect of the sub-filter scales has to be modelled. First, our results show that while the sharp-spectral filter has been shown to fulfil the needed mathematical properties \cite{aluie2009localness, eyink2009localness, speziale1985galilean, germano1986proposal, buzzicotti2018effect} and is the most obvious filter for pseudospectral simulations, it triggers wild fluctuations that blur the cascade-flux process and make difficult the understanding of its main properties. That seems related to the fact that most of quantities under investigation are found to be local in space, and hence our findings suggest to prefer the use of filters which are local and positive in physical space.

Then, our scale-by-scale analysis  shows that there is a strong correlation between the local energy flux rate and strain rate of the filtered  field. The conditional mean value of the flux rate follows a clear power-law dependence on the strain rate given by eq. (\ref{Smago}), making it predictable based on only filtered quantities. Thus the strain is a very good observable to characterize properties of the energy-flux. On the other hand the vorticity appears to be much more indirectly linked. The scaling relation observed in this work gives hence strong support to the Smagorinsky model and its variants \cite{smagorinsky1963general, pope2000turbulent} that use the strain to predict the sub-scale stress tensor. 
However, even though the Smagorinsky model predicts to good accuracy the mean values, it gives no prediction for the fluctuations around the mean that are of great importance as they control the inverse transfer events that are observed.

There are several steps that can be taken to extend this research. 
First of all, 
simulations at higher Reynolds number would be desirable in order to have a cleaner inertial range where the statistics are not affected neither by the forcing properties neither from viscous effects. 
Secondly, the present results were limited in considering correlations of the local flux with only the amplitude of the strain and the vorticity. This is a simplification that is required as a first step before examining more complex relations as the one given by he Clark model (\ref{clark}).
It was also noting that although the vorticity was not strongly correlating with the flux it was not completely disassociated from it. So a relation that involves both strain and vorticity is still a possible improvement of the Smagorinsky model. A fruitful direction that could be followed in future work is to examine the joint pdf of the energy flux and the invariants (under rotations and  reflections) of the gradient tensor (the so called QR \cite{chong1990general,ooi1999study}) that completely characterise the structure of the gradient tensor.  
Thirdly, it would be important of course to extend such an analysis to  bounded-flows, for instance a channel, which is key for applications.

Furthermore, even if an optimal parameterisation is chosen the energy flux will still depend on the sub-scale fluctuations that are essentially random in nature. One can not then hope to get an exact relation that connects the gradient tensor with the flux and this randomness will need to be taken in to account in terms of stochastic modeling. In terms of the Smagorinky model the simplest expression that generalises eq. (\ref{SmagoT}) can be given by
\be 
{\btau}_{i,j} \approx - C_s^2 \ell^2 S_\ell (1+ \xi_\ell) \overline{\bf S}_{i,j}
\label{SmagoSt}
\ee 
where $\xi_\ell$ is a zero mean spatio-temporal noise
that depends in principle on $\ell$ and $S_\ell$ and whose properties need to be determined from data.
Analysing the data displayed in figure \ref{fig7}, it turns out that it is not possible to fit all the curves with a simple random variable $\xi$, since non-trivial dependence of $\ell$ and $S_\ell$ is indeed found. Yet, the main features are decently recovered simply with a random variable whose pdf is given by $f(x)\propto \exp{(-\gamma_S\vert x \vert)}$, where $\gamma_S$ is a coefficient that has a dependence on $S$.
Finally, besides the energy the cascade of the second invariant that of helicity needs also to be studied, quantified and properly modeled. We plan to follow these directions in our future work.

\begin{acknowledgments}

This work was granted access to the HPC resources of MesoPSL financed by the R\'egion 
Ile de France and the project Equip@Meso (reference ANR-10-EQPX-29-01) of the programme Investissements d'Avenir supervised by the Agence Nationale pour la Recherche and the HPC resources of TGCC-CURIE \& CINES	Occigen (allocations No. A0070506421 \& No. A0062B10759) attributed by GENCI (Grand Equipement National de Calcul Intensif) where the present numerical simulations were performed. This work was also supported by the Agence nationale de la recherche (ANR DYSTURB project No. ANR-17-CE30-0004)

\end{acknowledgments}

\bibliographystyle{abbrv}
\bibliography{biblio}

\begin{thebibliography}{10}

\bibitem{alexakis2018cascades}
A.~Alexakis and L.~Biferale.
\newblock Cascades and transitions in turbulent flows.
\newblock {\em Physics Reports}, 767:1--101, 2018.

\bibitem{aluie2009localness}
H.~Aluie and G.~L. Eyink.
\newblock Localness of energy cascade in hydrodynamic turbulence. ii. sharp
  spectral filter.
\newblock {\em Physics of Fluids}, 21(11):115108, 2009.

\bibitem{Aumaitre:2001p6195}
S.~Auma{\^\i}tre, S.~Fauve, S.~McNamara, and P.~Poggi.
\newblock Power injected in dissipative systems and the fluctuation theorem.
\newblock {\em The European Physical Journal B}, Jan 2001.

\bibitem{bandi2009probability}
M.~Bandi, S.~G. Chumakov, and C.~Connaughton.
\newblock Probability distribution of power fluctuations in turbulence.
\newblock {\em Physical Review E}, 79(1):016309, 2009.

\bibitem{betchov1956inequality}
R.~Betchov.
\newblock An inequality concerning the production of vorticity in isotropic
  turbulence.
\newblock {\em Journal of Fluid Mechanics}, 1(5):497--504, 1956.

\bibitem{biskamp2003magnetohydrodynamic}
D.~Biskamp.
\newblock {\em Magnetohydrodynamic turbulence}.
\newblock Cambridge University Press, 2003.

\bibitem{bohr2005dynamical}
T.~Bohr, M.~H. Jensen, G.~Paladin, and A.~Vulpiani.
\newblock {\em Dynamical systems approach to turbulence}.
\newblock Cambridge University Press, 2005.

\bibitem{borue1998local}
V.~Borue and S.~A. Orszag.
\newblock Local energy flux and subgrid-scale statistics in three-dimensional
  turbulence.
\newblock {\em Journal of Fluid Mechanics}, 366:1--31, 1998.

\bibitem{buzzicotti2018effect}
M.~Buzzicotti, M.~Linkmann, H.~Aluie, L.~Biferale, J.~Brasseur, and
  C.~Meneveau.
\newblock Effect of filter type on the statistics of energy transfer between
  resolved and subfilter scales from a-priori analysis of direct numerical
  simulations of isotropic turbulence.
\newblock {\em Journal of Turbulence}, 19(2):167--197, 2018.

\bibitem{Chen:2006p1741}
J.~Chen, E.~Hawkes, R.~Sankaran, S.~Mason, and H.~Im.
\newblock Direct numerical simulation of ignition front propagation in a
  constant volume with temperature inhomogeneities i. fundamental analysis and
  diagnostics.
\newblock {\em Combustion and Flame}, 145(1-2):128--144, 2006.

\bibitem{chen2003joint}
Q.~Chen, S.~Chen, and G.~L. Eyink.
\newblock The joint cascade of energy and helicity in three-dimensional
  turbulence.
\newblock {\em Physics of Fluids}, 15(2):361--374, 2003.

\bibitem{chen2006kelvin}
S.~Chen, G.~L. Eyink, M.~Wan, and Z.~Xiao.
\newblock Is the kelvin theorem valid for high reynolds number turbulence?
\newblock {\em Physical review letters}, 97(14):144505, 2006.

\bibitem{chevillard2011lagrangian}
L.~Chevillard and C.~Meneveau.
\newblock Lagrangian time correlations of vorticity alignments in isotropic
  turbulence: Observations and model predictions.
\newblock {\em Physics of Fluids}, 23(10):101704, 2011.

\bibitem{Chevillard:2008p4119}
L.~Chevillard, C.~Meneveau, L.~Biferale, and F.~Toschi.
\newblock Modeling the pressure hessian and viscous laplacian in turbulence:
  comparisons with dns and implications on velocity gradient dynamics.
\newblock {\em Phys. Fluids}, 20(101504):1--15, 2008.

\bibitem{chong1990general}
M.~S. Chong, A.~E. Perry, and B.~J. Cantwell.
\newblock A general classification of three-dimensional flow fields.
\newblock {\em Physics of Fluids A: Fluid Dynamics}, 2(5):765--777, 1990.

\bibitem{ciliberto1998experimental}
S.~Ciliberto and C.~Laroche.
\newblock An experimental test of the gallavotti-cohen fluctuation theorem.
\newblock {\em Le Journal de Physique IV}, 8(PR6):Pr6--215, 1998.

\bibitem{domaradzki2007comparison}
J.~A. Domaradzki and D.~Carati.
\newblock A comparison of spectral sharp and smooth filters in the analysis of
  nonlinear interactions and energy transfer in turbulence.
\newblock {\em Physics of fluids}, 19(8):085111, 2007.

\bibitem{evans1993probability}
D.~J. Evans, E.~G.~D. Cohen, and G.~P. Morriss.
\newblock Probability of second law violations in shearing steady states.
\newblock {\em Physical review letters}, 71(15):2401, 1993.

\bibitem{Eyink:2006p1379}
G.~Eyink and K.~Sreenivasan.
\newblock Onsager and the theory of hydrodynamic turbulence.
\newblock {\em Reviews of Modern Physics}, Jan 2006.

\bibitem{eyink2009localness}
G.~L. Eyink and H.~Aluie.
\newblock Localness of energy cascade in hydrodynamic turbulence. i. smooth
  coarse graining.
\newblock {\em Physics of Fluids}, 21(11):115107, 2009.

\bibitem{falcon2008fluctuations}
E.~Falcon, S.~Auma{\^\i}tre, C.~Falc{\'o}n, C.~Laroche, and S.~Fauve.
\newblock Fluctuations of energy flux in wave turbulence.
\newblock {\em Physical review letters}, 100(6):064503, 2008.

\bibitem{Fox2003}
R.~O. Fox.
\newblock {\em Computational models for turbulent reacting flows}.
\newblock Cambridge Univ Press, 2003.

\bibitem{Fri_95}
U.~Frisch.
\newblock {\em Turbulence. {T}he legacy of {A.N} {K}olmogorov}.
\newblock Cambridge, University press, 1995.

\bibitem{gallavotti1995dynamical}
G.~Gallavotti and E.~G.~D. Cohen.
\newblock Dynamical ensembles in stationary states.
\newblock {\em Journal of Statistical Physics}, 80(5-6):931--970, 1995.

\bibitem{germano1986proposal}
M.~Germano.
\newblock A proposal for a redefinition of the turbulent stresses in the
  filtered navier--stokes equations.
\newblock {\em The Physics of fluids}, 29(7):2323--2324, 1986.

\bibitem{germano1992turbulence}
M.~Germano.
\newblock Turbulence: the filtering approach.
\newblock {\em Journal of Fluid Mechanics}, 238:325--336, 1992.

\bibitem{gill2016atmosphere}
A.~E. Gill.
\newblock {\em Atmosphere—ocean dynamics}.
\newblock Elsevier, 2016.

\bibitem{lesieur1996new}
M.~Lesieur and O.~Metais.
\newblock New trends in large-eddy simulations of turbulence.
\newblock {\em Annual review of fluid mechanics}, 28(1):45--82, 1996.

\bibitem{marconi2008fluctuation}
U.~M.~B. Marconi, A.~Puglisi, L.~Rondoni, and A.~Vulpiani.
\newblock Fluctuation--dissipation: response theory in statistical physics.
\newblock {\em Phys. Rep.}, 461(4-6):111--195, 2008.

\bibitem{meneveau2000scale}
C.~Meneveau and J.~Katz.
\newblock Scale-invariance and turbulence models for large-eddy simulation.
\newblock {\em Annual Review of Fluid Mechanics}, 32(1):1--32, 2000.

\bibitem{mininni2011hybrid}
P.~D. Mininni, D.~Rosenberg, R.~Reddy, and A.~Pouquet.
\newblock A hybrid mpi--openmp scheme for scalable parallel pseudospectral
  computations for fluid turbulence.
\newblock {\em Parallel Computing}, 37(6-7):316--326, 2011.

\bibitem{misra1997vortex}
A.~Misra and D.~I. Pullin.
\newblock A vortex-based subgrid stress model for large-eddy simulation.
\newblock {\em Physics of Fluids}, 9(8):2443--2454, 1997.

\bibitem{Moi_98}
P.~Moin and K.~Mahesh.
\newblock Direct numerical simulation: a tool in turbulence research.
\newblock {\em Ann. Rev. Fluid Mech.}, 30:539--578, 1998.

\bibitem{Mon_75}
A.~S. Monin and A.~M. Yaglom.
\newblock {\em Statistical {F}luid {M}echanics}.
\newblock MIT Press, Cambridge, Mass, 1975.

\bibitem{ooi1999study}
A.~Ooi, J.~Martin, J.~Soria, and M.~S. Chong.
\newblock A study of the evolution and characteristics of the invariants of the
  velocity-gradient tensor in isotropic turbulence.
\newblock {\em Journal of Fluid Mechanics}, 381:141--174, 1999.

\bibitem{piomelli1999large}
U.~Piomelli.
\newblock Large-eddy simulation: achievements and challenges.
\newblock {\em Progress in Aerospace Sciences}, 35(4):335--362, 1999.

\bibitem{Pope_turbulent}
S.~B. Pope.
\newblock {\em Turbulent Flows}.
\newblock Cambridge University Press, 2000.

\bibitem{pope2000turbulent}
S.~B. Pope.
\newblock {\em Turbulent Flows}.
\newblock Cambridge University Press, Cambridge, UK, 2000.

\bibitem{reynolds1990potential}
W.~C. Reynolds.
\newblock The potential and limitations of direct and large eddy simulations.
\newblock In {\em Whither turbulence? Turbulence at the crossroads}, pages
  313--343. Springer, 1990.

\bibitem{shang2005test}
X.-D. Shang, P.~Tong, and K.-Q. Xia.
\newblock Test of steady-state fluctuation theorem in turbulent
  rayleigh-b{\'e}nard convection.
\newblock {\em Physical Review E}, 72(1):015301, 2005.

\bibitem{smagorinsky1963general}
J.~Smagorinsky.
\newblock General circulation experiments with the primitive equations: I. the
  basic experiment.
\newblock {\em Monthly weather review}, 91(3):99--164, 1963.

\bibitem{speziale1985galilean}
C.~G. Speziale.
\newblock Galilean invariance of subgrid-scale stress models in the large-eddy
  simulation of turbulence.
\newblock {\em Journal of fluid mechanics}, 156:55--62, 1985.

\bibitem{Ten_90}
H.~Tennekes and J.~L. Lumley.
\newblock {\em A First Course in Turbulence}.
\newblock The MIT Press, Cambridge, Massachusetts, 1990.

\bibitem{thorpe2007introduction}
S.~A. Thorpe et~al.
\newblock {\em An introduction to ocean turbulence}, volume~10.
\newblock Cambridge University Press Cambridge, 2007.

\bibitem{tsinober2000vortex}
A.~Tsinober.
\newblock Vortex stretching versus production of strain/dissipation.
\newblock {\em Turbulence Structure and Vortex Dynamics}, pages 164--191, 2000.

\bibitem{tsinober2009informal}
A.~Tsinober.
\newblock {\em An informal conceptual introduction to turbulence}.
\newblock Springer, 2009.

\bibitem{vreman1997large}
B.~Vreman, B.~Geurts, and H.~Kuerten.
\newblock Large-eddy simulation of the turbulent mixing layer.
\newblock {\em Journal of fluid mechanics}, 339:357--390, 1997.

\bibitem{wilcox1998turbulence}
D.~C. Wilcox.
\newblock {\em Turbulence modeling for CFD}, volume~2.
\newblock DCW industries La Canada, CA, 1998.

\bibitem{zonta2016entropy}
F.~Zonta and S.~Chibbaro.
\newblock Entropy production and fluctuation relation in turbulent thermal
  convection.
\newblock {\em EPL (Europhysics Letters)}, 114(5):50011, 2016.

\end{thebibliography}

\end{document}